\documentclass[]{interact}
\usepackage{epstopdf}
\usepackage{xcolor}
\usepackage{soul}
\usepackage[caption=false]{subfig}

\usepackage[doublespacing]{setspace}

\usepackage[numbers,sort&compress]{natbib}
\bibpunct[, ]{[}{]}{,}{n}{,}{,}
\renewcommand\bibfont{\fontsize{10}{12}\selectfont}

\theoremstyle{plain}

\theoremstyle{definition}

\theoremstyle{remark}

\begin{document}

\articletype{ARTICLE TEMPLATE}

\title{Dislocation assisted coarsening of coherent precipitates: a phase field study}

\author{
\name{Arjun Varma R., Prita Pant and M. P. Gururajan}
\affil{Department of Metallurgical Engineering and Materials Science, Indian Institute of Technology Bombay, Powai, Mumbai, Maharashtra India 400076}
}

\maketitle

\begin{abstract}
Coarsening of precipitates in coherent systems is influenced by the elastic fields of the precipitates and the interfacial curvature. It is also known that if precipitates are connected by dislocations, coarsening is affected by the elastic fields of the dislocations and the pipe diffusivity. Although there is experimental evidence of accelerated coarsening in the presence of dislocations, these studies do not capture the effect of the elastic fields. There exist generic theoretical models that can predict the average sizes and size distributions of coarsening precipitates considering the coherency related elastic stress fields. In this paper, we use a phase field model to study the coarsening of precipitates connected by dislocations, incorporating its elastic fields and pipe diffusivity. Specifically, we study the effects of misfit strain, elastic moduli mismatch, faster pipe mobility and the elastic fields of the dislocation on the morphology and kinetics of the coarsening precipitates. The dilatational component, associated with the edge character of the dislocations interact with the precipitates in an elastically homogeneous system. In an elastically inhomogeneous system, the deviatoric elastic fields also interact with the precipitates influencing the morphology and kinetics of coarsening precipitates. The kinetics of coarsening is different as well, with hard precipitates coarsening faster as compared to soft precipitates, when connected by dislocations. Further, we note that the modified Gibbs-Thomson equation, which was originally derived for an isolated precipitate in an infinite matrix, is also applicable for coarsening precipitates at close proximity.
\end{abstract}


\section{Introduction}

The nucleation and growth of precipitates is followed by coarsening (or Ostwald ripening) during which larger precipitates grow at the expense of smaller precipitates; this is due to the Gibbs-Thomson effect which drives diffusion from regions of higher interfacial curvature to regions of lower interfacial curvature~\cite{Porter2009,Voorhees1992}. The size distribution and the average size of the precipitates during coarsening is given by the LSW (Lifshitz-Slyozov-Wagner) theory~\cite{Lifshitz1961,Wagner1961} albeit for the case when the volume fraction of the precipitates is very small. There exist modifications of this theory for finite volume fractions~\cite{Ardell1972,Brailsford1979,Voorhees1984,Marqusee1984,Enomoto1986}. 

In the case of coherent precipitates, the interfacial equilibrium is also influenced by the elastic fields of the precipitates. Hence, Johnson has proposed an addition to the Gibbs-Thomson equation~\cite{Johnson1987} to account for the effect of internal and external elastic fields of an isolated precipitate in an infinite system.  Thus, in the case of coherent precipitates, the kinetics of coarsening is altered~\cite{Johnson1984,Johnson1987,Abinandanan1993b}, and, at times, there is even inverse coarsening~\cite{Johnson1989,Johnson1990,Abinandanan1993b,Su1996a} where larger particles shrink and smaller particles grow. 

Experimentally, it is known that several systems such as (but not limited to) Al--Li~\cite{Schmitz1992,Wang1998}, Cu--Fe~\cite{Monzen2002,Shi2015}, Fe--Ti--C~\cite{Dunlop1975} and Fe--Cr~\cite{Kuchler2022} contain precipitates that remain coherent during coarsening, at least in the early stages. Dislocation activity in these systems result in dislocation glide towards the precipitates to minimise elastic energy and thereby connecting several precipitates. Such configurations may also be formed by the nucleation and growth of many precipitates on a dislocation line, which could act as a favourable site for precipitation~\cite{Dash1956,Bullough1959,Beaven1980,Kesternich1985,Wang1996}. 

When coherent precipitates and dislocations coexist, it is possible for the faster diffusivity of solutes along the dislocation core, known as ``pipe diffusion", to influence the coarsening behaviour of precipitates connected by them. There have been several experimental observations of accelerated coarsening of precipitates in the presence of dislocations~\cite{Dunlop1975,Allen1978,Zedalis1986,Angers1987,Taneike2001,Nakajima2004,Legros2008,Xia2016,Lin2016,Tkachev2020}. 
As opposed to the earlier studies which focused on an average effect of the dislocations on coarsening, the study by Legros et al~\cite{Legros2008} is important in that they have focused on just two precipitates connected by a dislocation. By attributing the dissolution of the smaller precipitate entirely to pipe diffusion, they also estimate that the pipe diffusivity in the Al--Si system is about three orders of magnitude higher than the lattice diffusivity. But, the thin-film geometry of the samples used in these experiments largely eliminate the effect of the elastic fields, hence missing out an important aspect of interaction of the coarsening precipitates with the dislocation. 

To summarise, there are theories that predict the average radius of particles and their size  distributions in systems with coherent and incoherent precipitates. However, these studies do not account for pipe diffusion, or account for them as average effects, and neglect the elastic interaction between the dislocations and precipitates. On the other hand, there are experiments on the effect of pipe diffusion on coarsening; but, in these studies, the geometry is such that there is no appreciable elastic interaction between dislocations and coherency stresses and strains. Hence, in this paper, we consider the coarsening of coherent precipitates in the presence of dislocations while accounting for both the elastic interactions and the faster solute diffusion along the dislocations -- albeit for the
case of two precipitates. Given the intensity of the computations, development of parallel codes 
is needed to study large number of coherent precipitates in the presence of dislocations. Having said that, as we show below, a careful study and a deeper understanding of
the effect of dislocations on coarsening of coherent precipitates is possible even in the
case of two precipitates.

Phase field models have been used extensively to study phase transformation phenomena~\cite{Chen2002} and deformation~\cite{Wang2010}. Recently, the effect of dislocations on phase transformations has also been studied using phase field models by evolving the composition field in presence of stationary dislocations, while incorporating the elastic fields and pipe diffusion along the dislocations~\cite{Varma2023}. In this paper, using the same approach, we explore the effect of dislocations on the coarsening behaviour of two spherical precipitates of different sizes connected by them. Specifically, we study the effects of misfit strain, elastic moduli mismatch, faster pipe mobility and the elastic fields of the dislocations on the morphology and kinetics of coarsening of two precipitates. We rationalise our simulation results in terms of the influence of the different parameters on the interfacial equilibrium, which,
in turn, alters the morphology and kinetics. 

This paper is organised as follows: in the next and following sections, we describe the phase-field model and the simulation details, respectively. In Sec.~\ref{Results} we present our results on coarsening kinetics, and particle morphologies and rationalise the same using Gibbs-Thomson and modified Gibbs-Thomson equations. Finally, we conclude the paper with a summary of some key conclusions.

\section{Phase-field model}

We use a phase field model capable of capturing the 
evolution of the two phases of a binary system and the dislocations, using the field variables $c$ and $\eta$. The model has been described in \cite{Varma2023} and is a combination of the phase 
field dislocation dynamics model~\cite{Lei2011,Hunter2011} 
and the Cahn-Hilliard model for composition evolution~\cite{Hu2001}. The field variables $c$ and $\eta$ 
are evolved according to the Cahn-Hilliard and 
Allen-Cahn kinetics respectively and the total 
free energy of the system is composed of chemical, 
interfacial, elastic and core energy components:
\begin{align}
F^{tot}(c,\eta) = &{N_V}\int \left(A^{ch}c^2(1-c)^2 + \kappa \left|\nabla c\right|^2 \right)\; dV \nonumber \\ &+ \frac{1}{2}\int 
C_{ijkl}(c)\epsilon^{el}_{ij}(c,\eta)\epsilon^{el}_{kl}(c,\eta)\; dV   + \sum_{\alpha=1}^N \int A^{core} sin^2(\pi\eta_{\alpha})\; dV
\label{tot-energy}
\end{align}
where $N_V$ is the number of atoms per unit volume, 
$A^{ch}$ is the bulk free energy coefficient that determines 
the height of the chemical energy barrier between the 
two phases, $\kappa$ is the coefficient of interfacial 
energy, $A^{core}$ is the coefficient of core energy 
and $V$ is the total volume of the system. As indicated in eq. 
(\ref{tot-energy}), the composition fields and the 
dislocations are coupled by their elastic fields; these elastic fields are, in turn, 
obtained by solving the equation of mechanical 
equilibrium,
\begin{equation} \label{MechEq}
\nabla \cdot \sigma^{el} = 0
\end{equation}
where $\sigma^{el}$ is the elastic stress.
Eq.~\eqref{MechEq} is solved using the Green's function approach~\cite{Mura1987}; we
assume linear elasticity and carry out homogenization using an iterative technique as described in~\cite{GuruThesis}. The modulus $C_{ijkl}$ is 
considered a function of 
the composition $c$ of the system, defined as: 
\begin{equation}
C_{ijkl}(c) = C^{\mathrm{0}}_{ijkl} + \alpha(c) \Delta C_{ijkl}
\end{equation} 
where, we have taken
$C^{\mathrm{0}}_{ijkl} = \frac{1}{2}\left[C^m_{ijkl}+C^p_{ijkl}\right]$ (an average of the matrix and precipitate moduli), $\Delta C_{ijkl} = C^{p}_{ijkl} - C^m_{ijkl}$ and $\alpha(c)$ is the interpolation function varying between 0 and 1 for $c=0$ to $c=1$, defined as: $\alpha(c) = \left[c^3\left(10 - 15c + 6c^2\right) -
\frac{1}{2}\right]$. The elastic fields for the system with the inhomogeneous moduli are obtained by an iterative method until the displacement fields converge, with mean error less than $10^{-8}$; see~\cite{GuruThesis} for details.

In this study, as we are interested in the 
morphological evolution of coherent precipitates and their coarsening (dissolution of the smaller precipitate and growth of the larger precipitate), we evolve only the composition field. 
The dislocations are considered stationary in our
simulations; the equilibrated  configuration of
dislocations is obtained by evolving 
the Allen-Cahn equation
in the absence of the composition field;  
these configurations are used
as the starting point for the coarsening
simulations.

The faster pipe mobility along the core of the dislocations is implemented according to the variable mobility formulation of Zhu et al.~\cite{Zhu1999}. The mobility $M$ for a system containing $N$ slip systems is given by:
\begin{equation}
M(\eta_{\alpha}) = M_{b} + M_{f}\left[\mathrm{max}\left(|\eta_1(1-\eta_1)|, |\eta_2(1-\eta_2)|,...,|\eta_N(1-\eta_N)|\right)\right]
\label{eq:pipe_mob}
\end{equation}
where $M_b=1$ is the bulk mobility and $M_f$ is the factor associated with the faster pipe mobility. In our simulations, we have taken $M_f=400$, such that the peak mobility at the dislocation $M_{\mathrm{peak}} = 101$. The pipe mobility term is highest when $\eta_{\alpha} = 0.5$, that is, at the dislocation core. 

All parameters used in our simulations are 
non-dimensionalised (see \cite{Varma2023} for 
details). The non-dimensionalised values of the 
key parameters of the model are shown in 
Table~\ref{tab:parameters}. We also consider two 
cases with an inhomogeneous elastic moduli; in 
those cased, the matrix moduli components are the 
same as that of the homogeneous, isotropic case 
in Table \ref{tab:parameters}, and the 
precipitate moduli components are taken as 20~\% 
higher and 20~\% lower than those of the matrix 
phase.

\begin{table}
    \tbl{Non-dimensionalised values of parameters used in our simulations}
    {\begin{tabular}{ccc} \toprule
       {\bf Parameter type} & {\bf Parameter} & 
       {\bf Value} \\ 
	\hline
        Simulation & $\Delta x=\Delta 
         y=\Delta z$ & 1.0 \\
        & $\delta t$ & 0.5 -- 1.0 \\
        Free energy & $A^{ch}$ & 1.0 \\
        & $A^{core}$ & 1.0 \\
        & $\kappa$ & 1.0\\
	Elastic & $\epsilon^c$ & 0.01 \\
	& $b$ & 1.0 \\     
        & $\frac{C_{1111}}{N_v}$ & 120 \\
	& $\frac{C_{1122}}{N_v}$ & 50 \\
	& $\frac{C_{1212}}{N_v}$ & 35 \\
        & Moduli contrast, $\delta$ & 0.8, 1.0, 1.2 \\
	& {Allowed error in displacements} & $10^{-8}$\\
    \hline
    \end{tabular}}
    \label{tab:parameters}
\end{table}

\section{Simulation details}
\begin{figure}[h]
    \centering
    \centering
	\subfloat{%
		\resizebox*{7cm}{!}{\includegraphics[width=9cm]{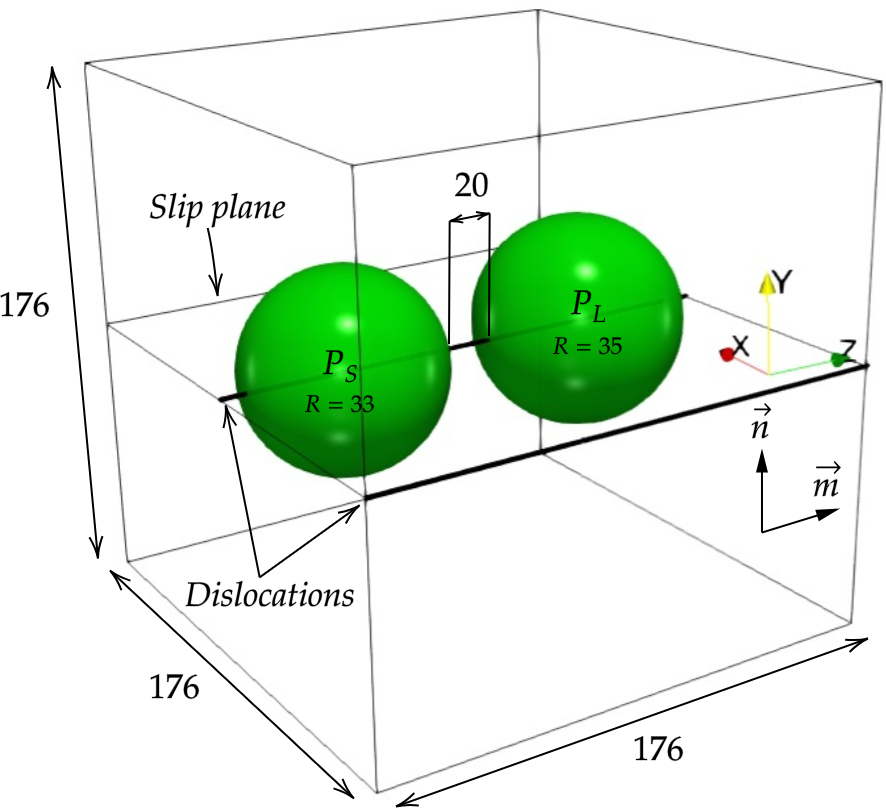}}}\hspace{5pt}
	%
    \caption{Simulation set-up consisting of two 
    spherical precipitates (shown by green c=0.5 
    iso-surface) connected by a dislocation loop 
    (shown in black). The slip plane in which the 
    dislocation exists is shown.The system 
    size is $176\times176\times176$. The directions 
    of the slip plane normal ($\Vec{n}$) and the sense vector ($\Vec{m})$ are 
    also indicated.}
    \label{fig:simulation-setup}
\end{figure}

We consider a system of $176 \times 176 
\times 176$ grid points; there are two spherical 
precipitates $P_S$ (smaller) and $P_L$ (larger) 
in this system which are connected by a 
straight dislocation 
line, as shown in 
Figure~\ref{fig:simulation-setup}. In our simulations, in order to be consistent with the periodic boundary conditions, we consider an infinitely long (along $z$ direction) straight dislocation dipole with one dislocation passing through the centre of the precipitates and the second dislocation situated on the xy face at a distance equal to half the simulation cell size: see Figure~\ref{fig:simulation-setup}.

We consider the following cases of simulations for 
our study: 
\begin{enumerate}
\item CC~(\textbf{C}oarsening due to interfacial \textbf{C}urvature) 
\item CCE~(\textbf{C}oarsening driven by interfacial \textbf{C}urvature and \textbf{E}lastic fields)
\item CCI~(\textbf{C}oarsening due to interfacial \textbf{C}urvature and elastic 
fields in an elastically \textbf{I}nhomogeneous system) 
\item CCED (\textbf{C}oarsening driven by interfacial \textbf{C}urvature and \textbf{E}lastic fields in the presence of a connecting 
\textbf{D}islocation). The dislocations affect
the coarsening in the following two ways:
\begin{enumerate}
\item The mobility is faster (pipe mobility) along the dislocation line.
\item The elastic fields of the dislocation interact with those of the solutes.
\end{enumerate}
\item CCID~(\textbf{C}oarsening due to 
interfacial \textbf{C}urvature and elastic 
fields in an elastically \textbf{I}nhomogeneous 
system in the presence of \textbf{D}islocations) 
\end{enumerate}

We consider isotropic elastic moduli for both the 
precipitate and the matrix in all cases. The initial 
radii of $P_S$ and $P_L$ are fixed at 33 and 35 
non-dimensionalised length units (for the small and 
large precipitate, respectively) and the surface to 
surface distance (measured along the z direction) is 
equal to 20 units. In order for the precipitates to be 
stable and not shrink, we need to maintain corresponding 
supersaturation in the matrix. In all cases, we maintain 
a supersaturation that will support the smaller precipitate. 
However, this supersaturation value will make the bigger 
precipitate to grow. Hence, the difference in radii between 
the two precipitates is kept to a minimum to ensure that 
there is negligible amount of supersaturation for $P_L$ 
and the increase in its volume occurs primarily at the 
expense of $P_S$. The results for the CC and the CCE cases 
are presented from t=10000 onwards in order to avoid the 
initial transients and compositional adjustments. The 
composition profile at t=10000 from the CCE (CCI) case is 
used as the initial composition profile in the case of 
the CCED (CCID) simulations. 

\section{Results and discussion}\label{Results}

In all the cases, we present the precipitate volume fraction as a function of time (which stays
nearly a constant since we are in the coarsening 
regime), the change in the individual precipitate sizes with time ($P_S$ shrinks while $P_L$ grows),
and the morphological changes, if any; in all 
cases, we also show and discuss the Gibbs-Thomson
effect. We use the Case CC as the baseline for
discussing all the subsequent simulation results.

\subsection{Case CC: Effect of interfacial curvature}

We begin the simulations with a far-field composition $c_{f} = 0.0101$, since this is the equilibrium composition at the interface corresponding to the spherical precipitate $P_S$ of radius 33 units, obtained using the Gibbs-Thomson equation~\cite{Johnson1987} (see Supplementary information for calculation):
\begin{equation}
	\Delta c^m_{int} = \frac{2\gamma 
		\chi^p}{(c^p_{\infty}-c^m_{\infty})\Psi_m}
	\label{eq:curvature-effect}
\end{equation}
where $\gamma$ is the interfacial energy of the 
system, $\chi^p$ is the mean curvature of the 
precipitate with respect to the precipitate 
phase, and $c^p_{\infty}$, $c^m_{\infty}$ are the equilibrium 
compositions of the matrix (zero for the double well potential assumed by us) and precipitate (unity for the double well potential assumed by us) phases respectively when separated by a flat interface (that is, $R \rightarrow \infty$). $\Psi_m$ is the curvature of the bulk free energy 
density versus composition curve, calculated at $c^m_{\infty}$. Note that the interfacial equilibrium compositions at the matrix and precipitate sides of the curved interface $\Delta c^m$ and $\Delta c^p$ are also equal for the double well potential, as the curvatures of bulk free energy densities $\Psi_m$ and $\Psi_p$, calculated at $c^m_{\infty}$ and $c^p_{\infty}$ respectively, are equal. In the same way, we take the initial compositions of the precipitates $P_S$ and $P_L$ as 1.0101 and 1.0095, respectively.

\begin{figure}
	\centering
	\subfloat[]{%
		\resizebox*{7cm}{!}{\includegraphics[width=9cm]{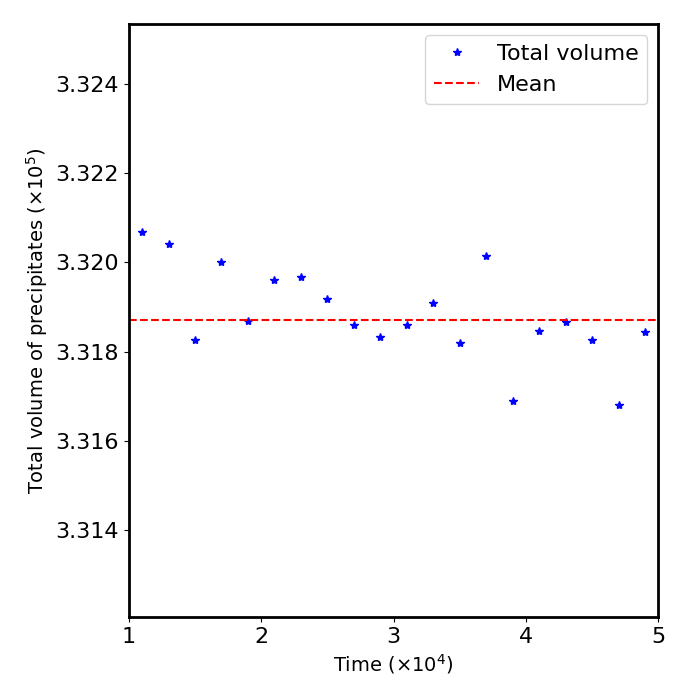}}}\hspace{5pt}
	\subfloat[]{%
		\resizebox*{7cm}{!}{\includegraphics[width=9cm]{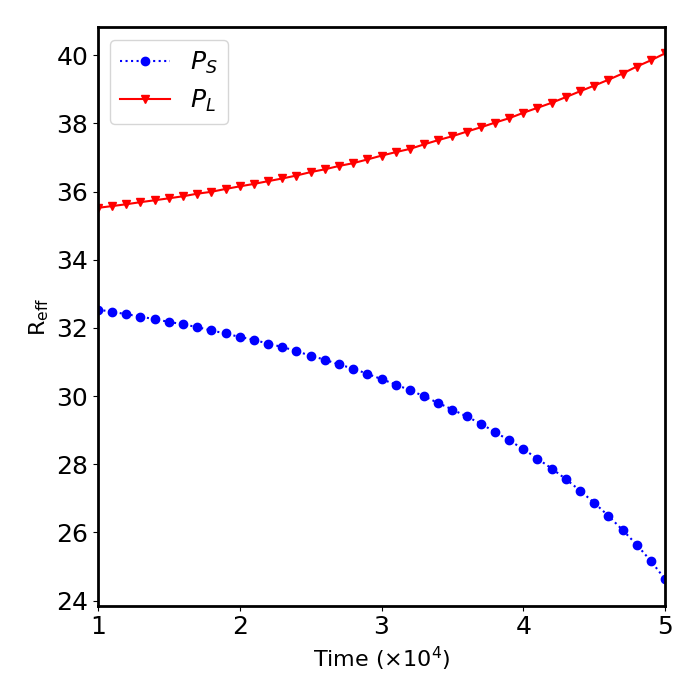}}}
\caption{Variation of (a) total volume of precipitates with time and (b) $\mathrm{R}_{\mathrm{eff}}$ of smaller ($P_S$) and larger ($P_L$) precipitates with time. Both profiles are shown only from t=10000, after the initial adjustments. } \label{fig:CC_rad-vol}
\end{figure}

In Figure~\ref{fig:CC_rad-vol}, we show the the total 
volume of the precipitates and the effective radius 
of the two precipitates as a function of time. The 
effective radii of the precipitates are calculated as:
\begin{equation}
	R_{\mathrm{eff}} = \left({\frac{3 V}{4\pi}}\right)^{\frac{1}{3}}
\end{equation}
where $V$ is the volume of the precipitate, 
obtained by counting the number of grid points 
with $c>=0.5$. The total volume of the precipitate remains 
constant throughout the simulation; note that
the oscillations seen are an artifact of the way
in which the volume fraction is estimated. The effective 
radius of the smaller precipitate decreases and the larger 
precipitate increases. 
Hence, it is clear that the 
precipitate size changes are due to coarsening. 

\begin{figure}
	\centering
	\subfloat[]{%
		\resizebox*{7cm}{!}{\includegraphics[width=9cm]{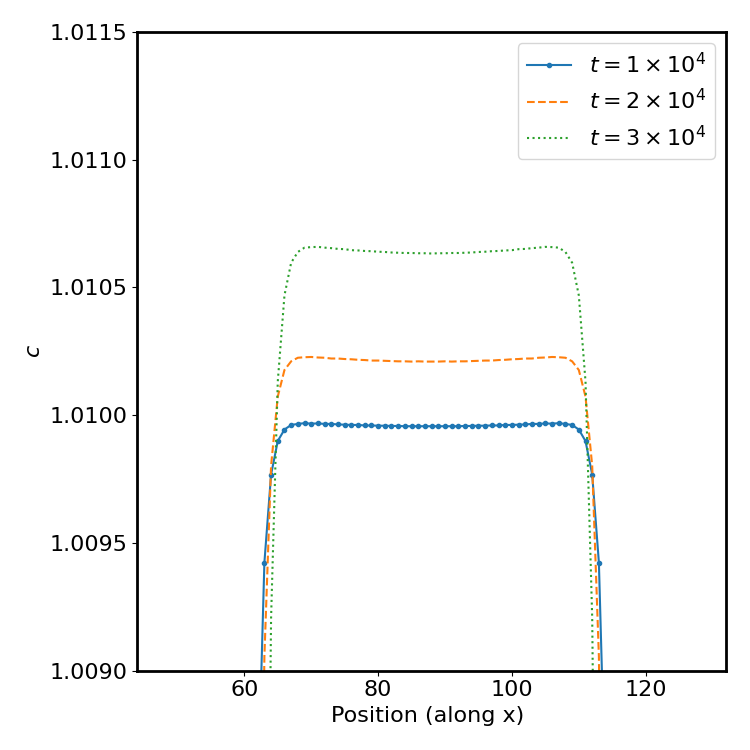}}}\hspace{5pt}
	\subfloat[]{%
		\resizebox*{7cm}{!}{\includegraphics[width=9cm]{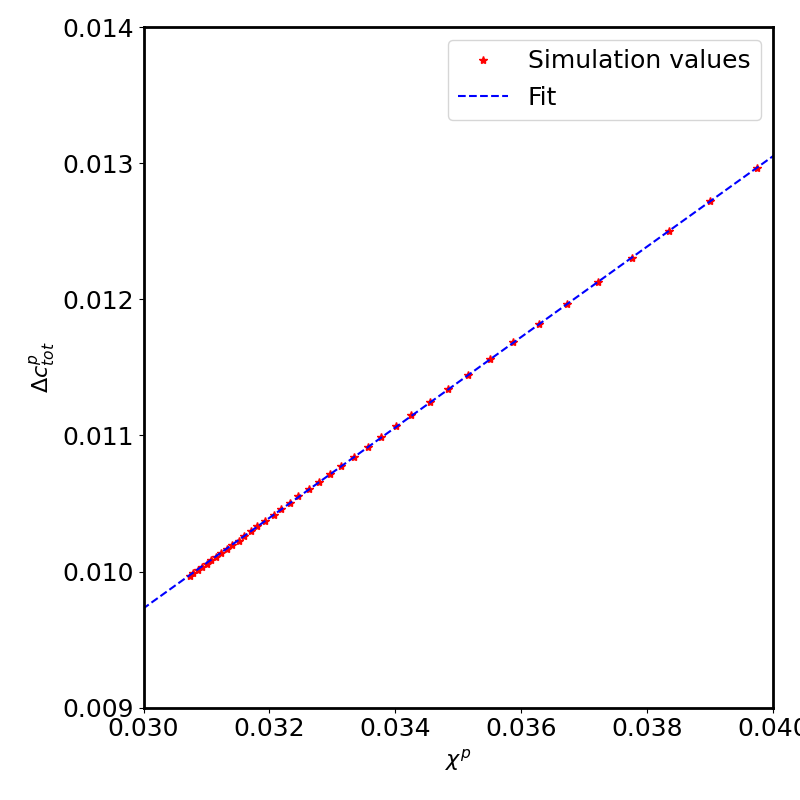}}}
	\caption{(a) Composition profile of smaller precipitate $P_S$ (plotted along the x-direction) increases as $\mathrm{R}_{\mathrm{eff}}$ decreases. (b) The fit of $\Delta c^p_{int} = c^p_{R}-c^p_{\infty}$ against mean curvature $\chi^p$ (dotted line) superimposed on the values obtained from the simulations. The fit has $R^2=1$ and a residual standard error of $4.49\times10^{-6}$. The standard errors associated with the slope and intercept are respectively equal to $2.544\times10^{-4}$ and $8.592\times10^{-6}$.} \label{fig:CC_comp_profile_deltacpfit}
\end{figure}

The composition profile of the smaller precipitate ($P_S$) is plotted at different stages of evolution in Figure~\ref{fig:CC_comp_profile_deltacpfit} (a). Due to Gibbs-Thomson effect, the equilibrium composition inside the precipitate $c^p_R$ increases as the $\mathrm{R}_{\mathrm{eff}}$ of $P_S$ decreases. In Figure~\ref{fig:CC_comp_profile_deltacpfit} (b),
the $\Delta c^p_{int}=c^p_{R}-c^p_{\infty}$ at $P_S$ is plotted against the inverse of $\mathrm{R}_{\mathrm{eff}}$; since the precipitate
remains spherical, inverse of $\mathrm{R}_{\mathrm{eff}}$  is nothing but the mean curvature ($\chi^p$) of the precipitate-matrix interface. According to equation~(\ref{eq:curvature-effect}), a linear fit of $\Delta c^p_{int}$ against $\chi^p$ should yield a slope equal to the interfacial energy $\gamma$ and a zero intercept. 
The slope obtained from our fit is 0.33, which is 
close to the interfacial energy 
$\gamma=\frac{1}{3}$ for our choice of $A^{ch}$ 
and $\kappa$. Further, the intercept of the 
fitted line is equal to -0.0002, close to the 
expected value of zero. The same exercise was 
also repeated for the precipitate $P_L$ for which 
the slope and intercept were respectively equal 
to 0.33 and -0.0004. The agreement of the values 
obtained from the fit with the theoretically 
calculated values serves as a benchmark for our 
code. 

\subsection{Case CCE: Effect of precipitate coherency}

\begin{figure}[h]
	\centering
	\subfloat[]{%
		\resizebox*{7cm}{!}{\includegraphics[width=9cm]{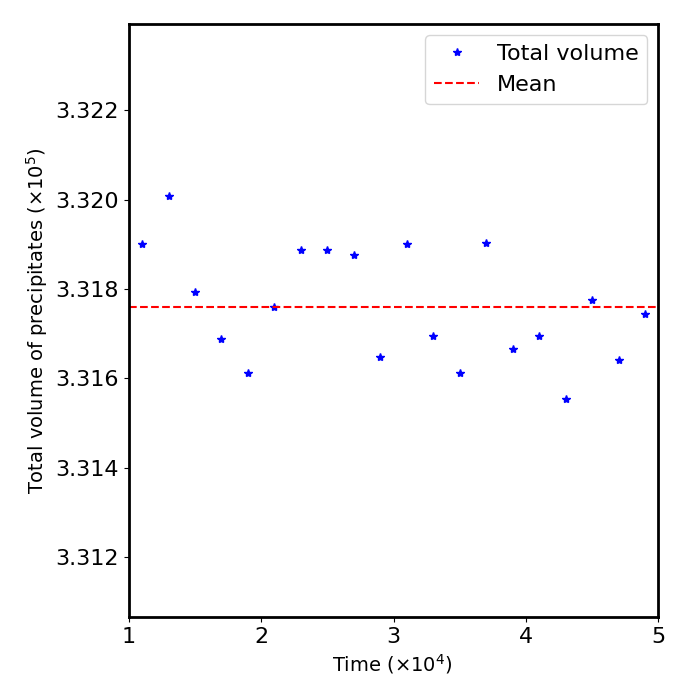}}}\hspace{5pt}
	\subfloat[]{%
		\resizebox*{7cm}{!}{\includegraphics[width=9cm]{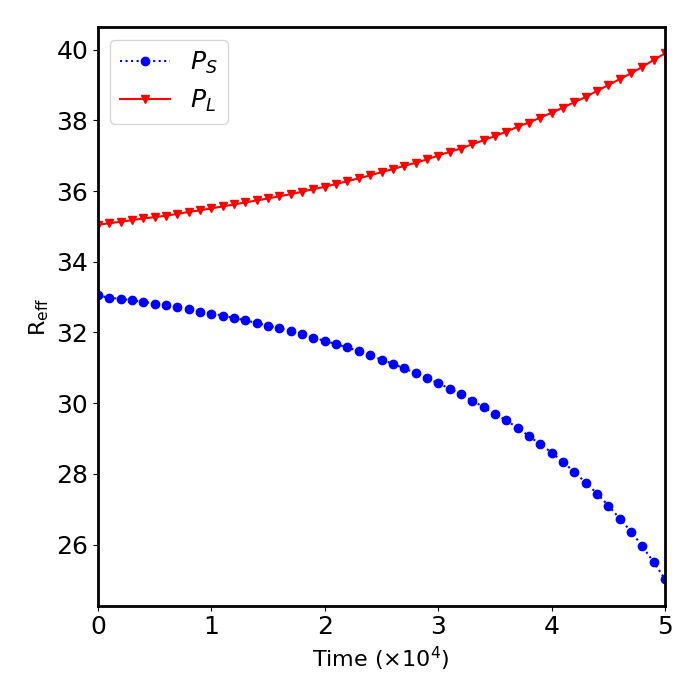}}}
	\caption{Variation of (a) total volume of precipitates with time (red line is the mean) and (b) $\mathrm{R}_{\mathrm{eff}}$ of smaller ($P_S$) and larger ($P_L$) precipitates with time for the CCE case.} \label{fig:CCE_rad-vol}
\end{figure}
 
In these simulations, the far-field composition is taken to be $c_f= 0.0165$. This is equal to the equilibrium interfacial composition for the smaller precipitate ($P_S$ of radius 33) when the magnitude of misfit strain $\epsilon^c=0.01$. This is calculated according to the modified Gibbs-Thomson equation, which incorporates 
the effect of elastic fields of the precipitates~\cite{Johnson1987} (the second term
in the following equation):
\begin{equation}
\Delta c^m_{tot} = \frac{2\gamma 
\chi^p}{(c^p_\infty-c^m_\infty)\Psi_m} + 
\frac{\sigma^m_{ij}(\epsilon^m_{ij}-\epsilon^p_{ij})
 + 
\frac{1}{2}\left[\sigma^p_{ij}(\epsilon^p_{ij}-\epsilon^c\delta_{ij})
-\sigma^m_{ij}\epsilon^m_{ij}\right]}{(c^p_\infty-c^m_\infty)\Psi_m}
\label{eq:modified_GT}
\end{equation} 
where $\sigma^m, \;\sigma^p$ and $\epsilon^m,\;\epsilon^p$ are the stresses and total strains at the matrix and precipitate sides of the interface and $\epsilon^c$ is the magnitude of the misfit strain of the precipitate. (see Supplementary information for calculation.)

\begin{figure}[h]
	\centering
	\subfloat[]{%
		\resizebox*{7cm}{!}{\includegraphics[width=9cm]{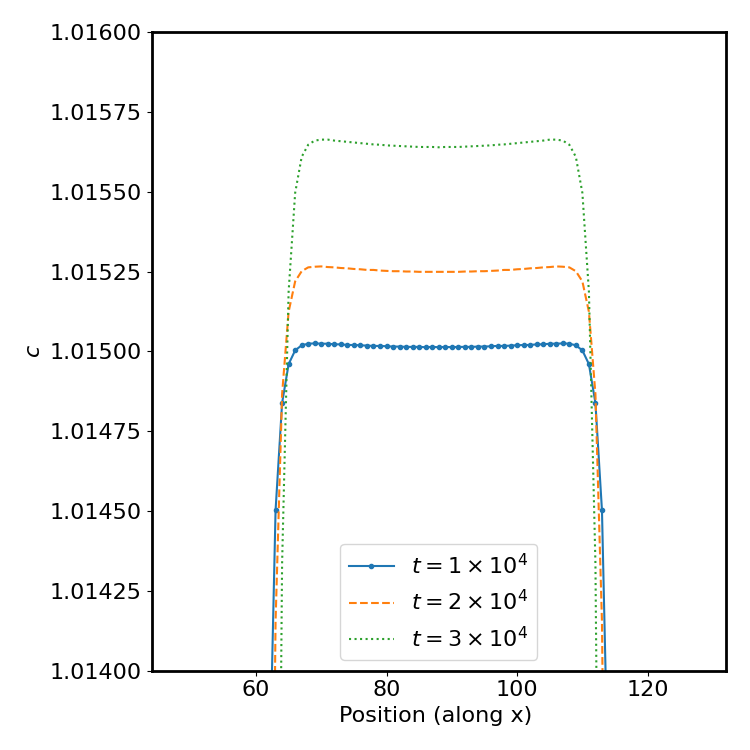}}}\hspace{5pt}
	\subfloat[]{%
		\resizebox*{7cm}{!}{\includegraphics[width=9cm]{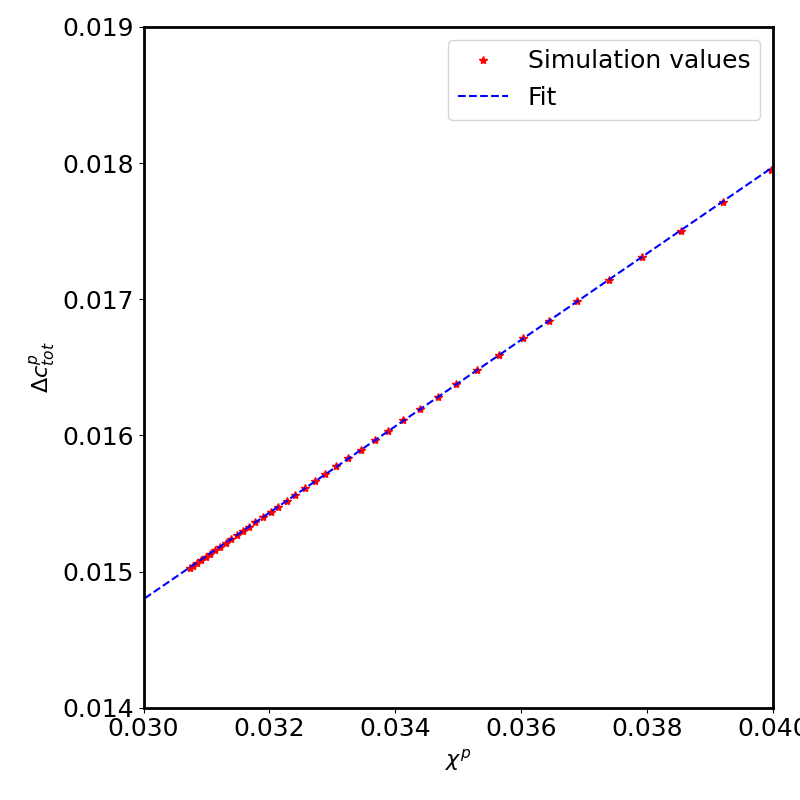}}}
    \caption{(a) Composition profiles plotted along the $x$ direction of smaller precipitate at different timesteps shows an increase in equilibrium composition of precipitate with a decrease in size of the precipitate. (b) A linear fit of $\Delta c^p_{tot}=c^p_R-c^p_{\infty}$ against mean curvature $\chi^p=\frac{1}{\mathrm{R}_{\mathrm{eff}}}$ produces a y-intercept of 0.0053. The fit has $R^2=1$ and a residual standard error of $3.51\times10^{-6}$. The standard errors associated with the slope and intercept of the fit are respectively $2.178\times10^{-4}$ and $7.321\times10^{-6}$.} 
   \label{fig:CCE_comp_profile_deltacpfit}
\end{figure}

Figure~\ref{fig:CCE_rad-vol} shows the evolution of the total volume of the two precipitates and the radius of each precipitate respectively, clearly showing that the system is within the coarsening regime. In Figure~\ref{fig:CCE_comp_profile_deltacpfit}, we show the equilibrium composition inside the smaller precipitate ($P_S$) at different time steps time and a linear fit of $\Delta c^p_{tot}=c^p_R-c^p_\infty$ against the mean curvature $\chi^p$ (=$\frac{1}{\mathrm{R}_{\mathrm{eff}}}$). As the radius of the smaller precipitate $P_S$ decreases, the equilibrium composition within the precipitate ($c^p_R$) becomes higher. The linear fit yields a slope equal to 0.32 and a $y$-intercept equal to 0.0053. In the CCE case, the $y$-intercept is equal to the second term in eq.~(\ref{eq:modified_GT}), that is the elastic contribution to $\Delta c^p_{tot}$. The elastic contribution $\Delta c^p_{el}$ may also be calculated by obtaining the elastic fields of the precipitate by solving Eshelby's inclusion problem~\cite{Mura1987}. The value of $\Delta c^p_{el}$ obtained using this approach is equal to 0.0064. We believe that this difference in intercepts is due to the elastic field overlap of the two precipitates and hence the Eshelby solution which is meant for the isolated precipitate has to be corrected by including this elastic interaction.

\begin{figure}[h]
	\centering
	\subfloat{%
		\resizebox*{8cm}{!}{\includegraphics[width=9.5cm]{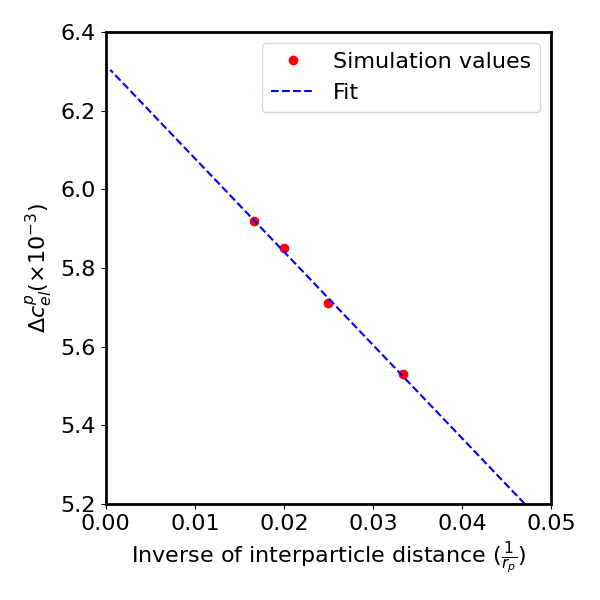}}}\hspace{5pt}
	\caption{Finite-size scaling of $\Delta c^p_{el}$ against the inverse of inter-particle separation ($\frac{1}{r_p}$). The inter-particle separations of 30, 40, 50 and 60 units are obtained in grids of size $196^3$, $216^3$, $236^3$ and $256^3$ respectively. The $y$-intercept (corresponding to a $\Delta c^p_{el}$ at infinite interparticle spacing) is 0.0063.  The fit has $R^2=0.9963$ and a residual standard error of $1.252\times10^{-5}$. The standard errors associated with the slope and the intercept of the fitted line are respectively $9.94\times10^{-5}$ and $2.45\times10^{-5}$.} \label{fig:finite-size-scaling}
\end{figure}

Hence, in order to rationalise this difference between the value obtained from 
the simulation and theory, we carried out a finite-size scaling analysis by varying the interparticle distance. Figure~\ref{fig:finite-size-scaling} shows the $\Delta c^p_{el}$ contribution at different interparticle spacings. There is an increase in the value of $\Delta c^p_{el}$ as the interparticle 
spacing increases. A linear fit of $\Delta c^p_{el}$ values against 
the inverse of interparticle spacing ($\frac{1}{r_p}$) yields a 
y-intercept of 0.0062, close to the theoretically calculated value. Hence, the difference in $\Delta c^p_{el}$ from our simulations with respect to the theoretical calculation may be attributed to the interaction between the precipitates. 

It is also worth noting that the elastic fields of a precipitate are independent of the size of the precipitate~\cite{Johnson1987}. Hence, the $\Delta c^p_{el}$ contribtions for both precipitates 
remain the same in the case of CCE simulations. Hence, the concentration gradient between the two precipitates should remain the same for CC and CCE cases, resulting in same kinetics of evolution. The difference in evolution kinetics between the two cases is minimal, as expected, and the minor difference is due to the slightly larger supersaturation for smaller precipitate ($P_S$) in the CCE case $(c_{f}=0.0165)$, which was calculated by taking $\Delta c^p_{el}$ equal to the theoretically calculated value of 0.0064 (see Supplementary information).

\subsection{Case CCED (a): Effect of pipe diffusivity of dislocation}

Next, we consider the effect of dislocations, but only due to the faster pipe diffusion along the dislocation core. In the study by Legros et al~\cite{Legros2008}, the dissolution of the smaller precipitate was assumed to occur only by virtue of pipe diffusion through the dislocation. However, for our simulations, such an assumption would be incorrect. Hence, a direct comparison of our results with the experiment is not possible, as diffusion occurs 
both through the dislocation core and the bulk. So, we calculate 
the apparent mobility of the system due to the introduction of the 
dislocations in the system. 

\begin{figure}[h]
	\centering
	\subfloat[]{%
		\resizebox*{7cm}{!}{\includegraphics[width=8cm]{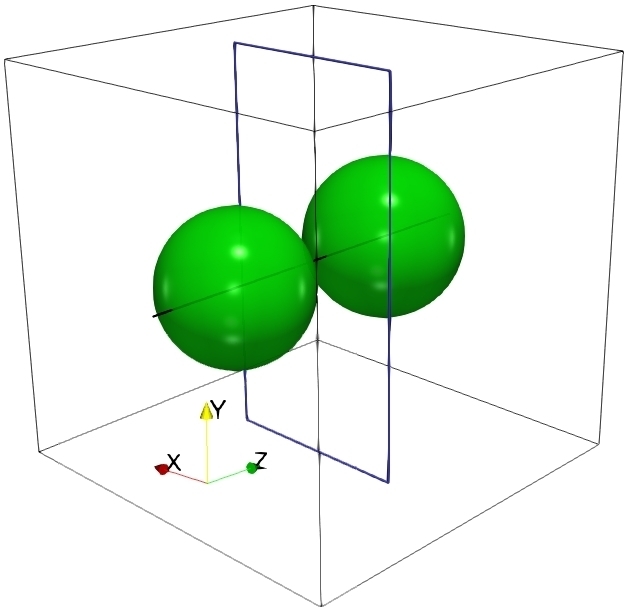}}}\hspace{5pt}
        \subfloat[]{%
		\resizebox*{7cm}{!}{\includegraphics[width=5cm]{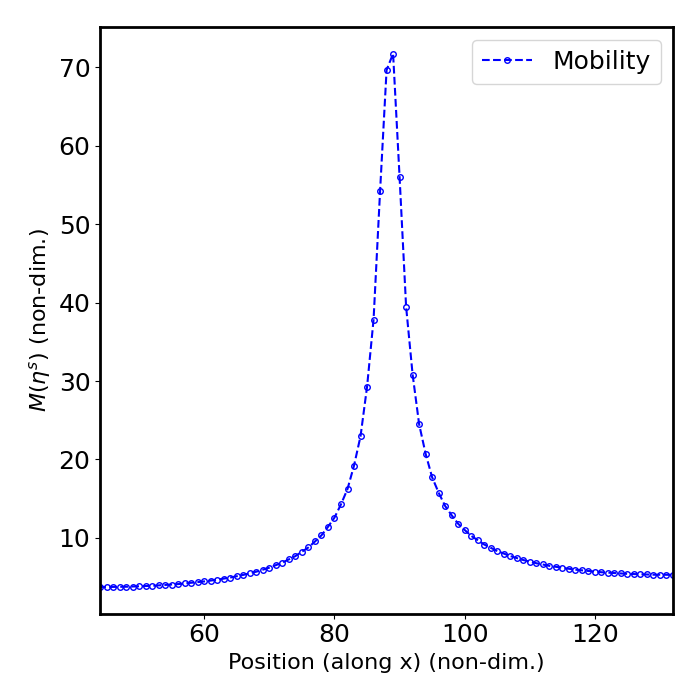}}}\hspace{5pt}
	\caption{Schematic of (a) the $xy$ plane 
	between the two precipitates used for 
	calculating the area fraction of the pipe and 
	the flux along the $z$-direction. (b) 
	Position dependent mobility plotted along a 
	line taken in the slip plane and 
	perpendicular to the line direction of the 
	dislocations. (See supplementary information 
	for a 2D map of position dependent mobility 
	perpendicular to the slip plane.)} 
	\label{fig:area_fraction_schematic}
\end{figure}

In order to obtain the apparent atomic mobility, we follow the definition of apparent diffusivity in~\cite{Porter2009}:
\begin{equation}
D_{\mathrm{app}} = D_{\mathrm{lattice}} + a_f D_{\mathrm{pipe}}
\label{eq:app_mob}
\end{equation}
where $D_{\mathrm{lattice}}$ is the diffusivity in the defect free lattice, $a_f$ is the area fraction and $D_{\mathrm{pipe}}$ is the diffusivity through the dislocation pipe. By definition, the atomic mobility and diffusivity are related as $D=M \Psi$, where $\Psi=\frac{\partial^2 f}{\partial c^2}$ is the curvature of the double well potential. Hence, eq.~(\ref{eq:app_mob}) will approximate for atomic mobility as well, as $\Psi$ is the same on both sides and hence can be cancelled  out. In our simulations, the pipe mobility is defined according to eq.~(\ref{eq:pipe_mob}), which peaks at the dislocation line and drops down to the bulk mobility value in a diffuse manner (see supplementary information). For the choice of our parameters, the peak value of mobility is equal to $M_{\mathrm{peak}}=101$, two orders of magnitude higher than the mobility in the bulk. However, as the mobility is varying in a diffuse manner between these values, we have adopted the following definitions for $M_{\mathrm{lattice}}$, $M_{\mathrm{pipe}}$, and $a_f$. We consider a plane perpendicular to the slip plane, which is also perpendicular to the dislocation line, as shown in Figure~\ref{fig:area_fraction_schematic}. 

Next, we define the dislocation 
pipe to be the grid points on this plane, where $M\geq \frac{M_{\mathrm{peak}}}{10}$ and the grid points where $M<\frac{M_{\mathrm{peak}}}{10}$, as the bulk region. The average mobility of all points in the pipe region was taken as $M_{\mathrm{pipe}}$ and that 
of all points outside of this region was taken as $M_{\mathrm{lattice}}$. The area fraction of the pipe was calculated as the fraction of grid points on the plane within the pipe region. The value $\frac{M_{\mathrm{peak}}}{10}$ is chosen to define the pipe region as the mobility outside this region may be considered almost invariant. 

Using this method for calculation, we obtain $M_{\mathrm{lattice}}=2.031$, $M_{\mathrm{pipe}}=17.91$ and $a_f=0.00464$. Hence, the apparent mobility of the system for our choice of parameters is:
\begin{align}
M_{\mathrm{app}} &= M_{\mathrm{lattice}} + a_f M_{\mathrm{pipe}} \nonumber\\
                &= 2.031 + 0.00464\times17.91 \\ 
                &=2.114 \nonumber
\end{align}

It is also possible to calculate $M_{\mathrm{app}}$ by taking the ratio of the flux across the plane in Figure~\ref{fig:area_fraction_schematic} from the smaller to the larger precipitate in the $z$-direction. This is because the 
pipe diffusion does not alter the energetics of the system and hence the driving force for diffusion remains the same in both cases. The flux is calculated as:
\begin{equation}
J_z = -N_v M \left[\frac{\partial \mu^{tot}}{\partial z}\right]
\end{equation}
where $M$ is the atomic mobility of the system (position dependent for the case CCED (a)), $\frac{\partial \mu^{tot}}{\partial z}$ is the gradient in the total potential along the $z$-direction, where $\mu^{tot}=\left(\frac{1}{N_v}\frac{\delta F^{tot}(c,\eta^{\alpha})}{\delta c}\right)$ is the variational derivative of the total free energy of the system with respect to composition. Hence, the ratio of the fluxes 
calculated in the CCE and CCED (a) case is expected to give the apparent mobility $M_{\mathrm{app}}$ (because $M=1$ in CCE case). Table~\ref{tab:flux_ratio} shows the fluxes calculated for the two simulation cases and the apparent mobility, obtained as the ratio of the two fluxes, at different effective radii of the smaller precipitate $P_S$. 

\begin{table}[h!]
	\centering
	\tbl{Apparent mobility of CCED(a) case keeping CCE case as the reference.}
	{\begin{tabular}{cccc} 
			\hline
			{\bf $\mathrm{R}_{\mathrm{eff}}$ of 
			$P_S$} & 
			{\bf 
				$J_z$ 
				(CCE)} &  {\bf $J_z$ (CCED)} & 
			$M_{\mathrm{app}}$ \\ 
			\hline
			32 & 0.46 & 1.01 & 2.19 \\
			30 & 0.80 & 1.80 & 2.25 \\
			28 & 1.10 & 2.41 & 2.19 \\
			26 & 1.35 & 2.92 & 2.16 \\
			\hline
	\end{tabular}}
	\label{tab:flux_ratio}
\end{table} 

The $M_{\mathrm{app}}$ value obtained by taking the ratio of the fluxes from the two simulation cases agree well with the value obtained from the calculation based on eq.~(\ref{eq:app_mob}), indicating that the accelerated diffusion kinetics along the dislocation is correctly captured in our simulations.

\subsection{Case CCED (b): Effect of elastic fields of dislocation (along with pipe diffusivity)}

\begin{figure}[h!]
\centering
\subfloat{%
\resizebox*{13cm}{!}{\includegraphics[width=9cm]{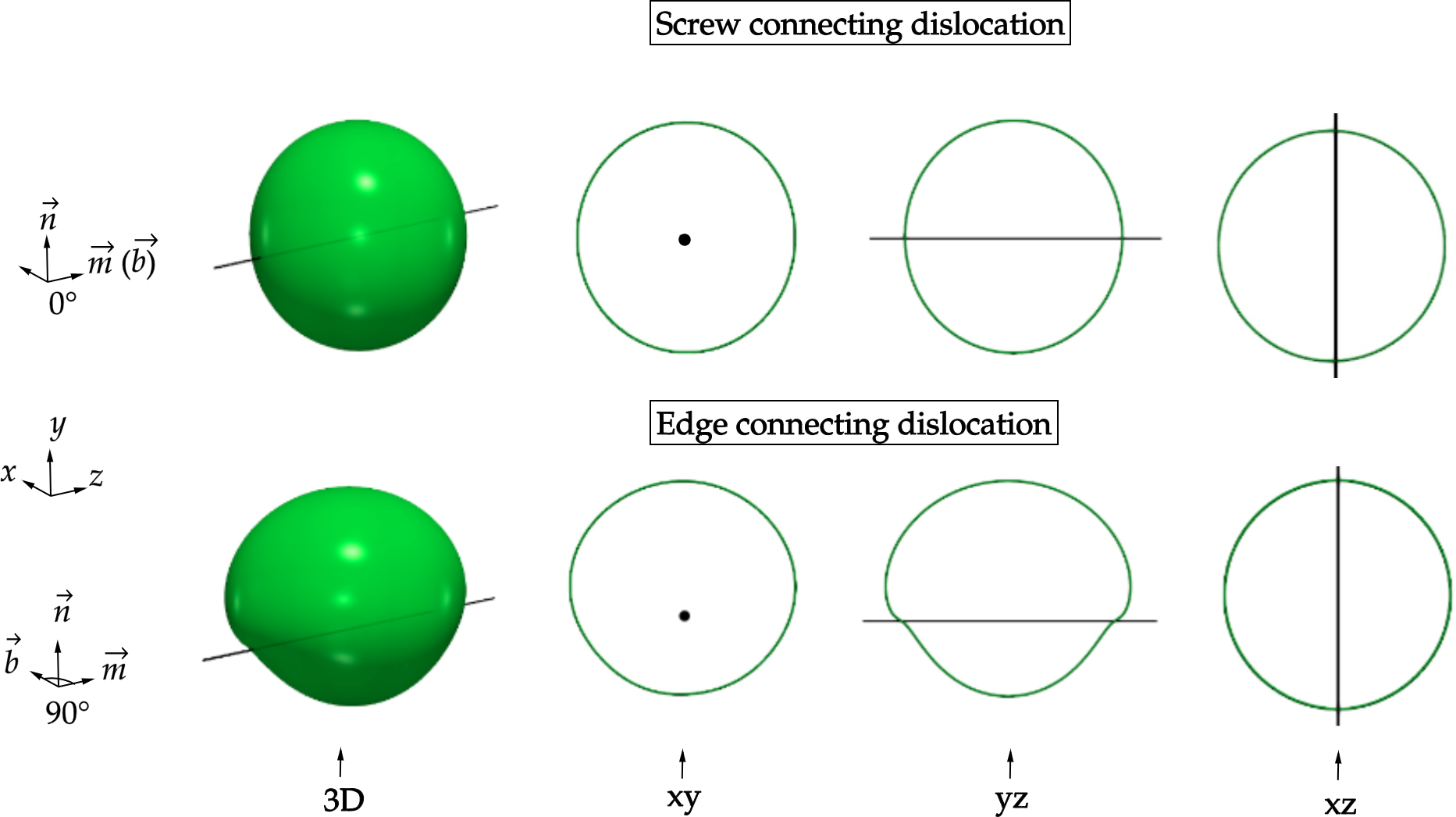}}}\hspace{5pt}
\caption{Morphology of smaller precipitate $P_S$ connected by screw (first row) and edge (second row) dislocations at t=12500}
\label{fig:edge_screw_morphology}
\end{figure}

Next, we consider the results from the CCED case incorporating the strain fields of the dislocations. In Figure~\ref{fig:edge_screw_morphology}, we show the $xy$, $yz$ and $xz$ sections of the smaller precipitate ($P_S$) at time t=12500 for the screw and edge type connecting dislocations respectively.

\begin{figure}[h!]
\centering
\subfloat{%
\resizebox*{12cm}{!}{\includegraphics[width=10cm]{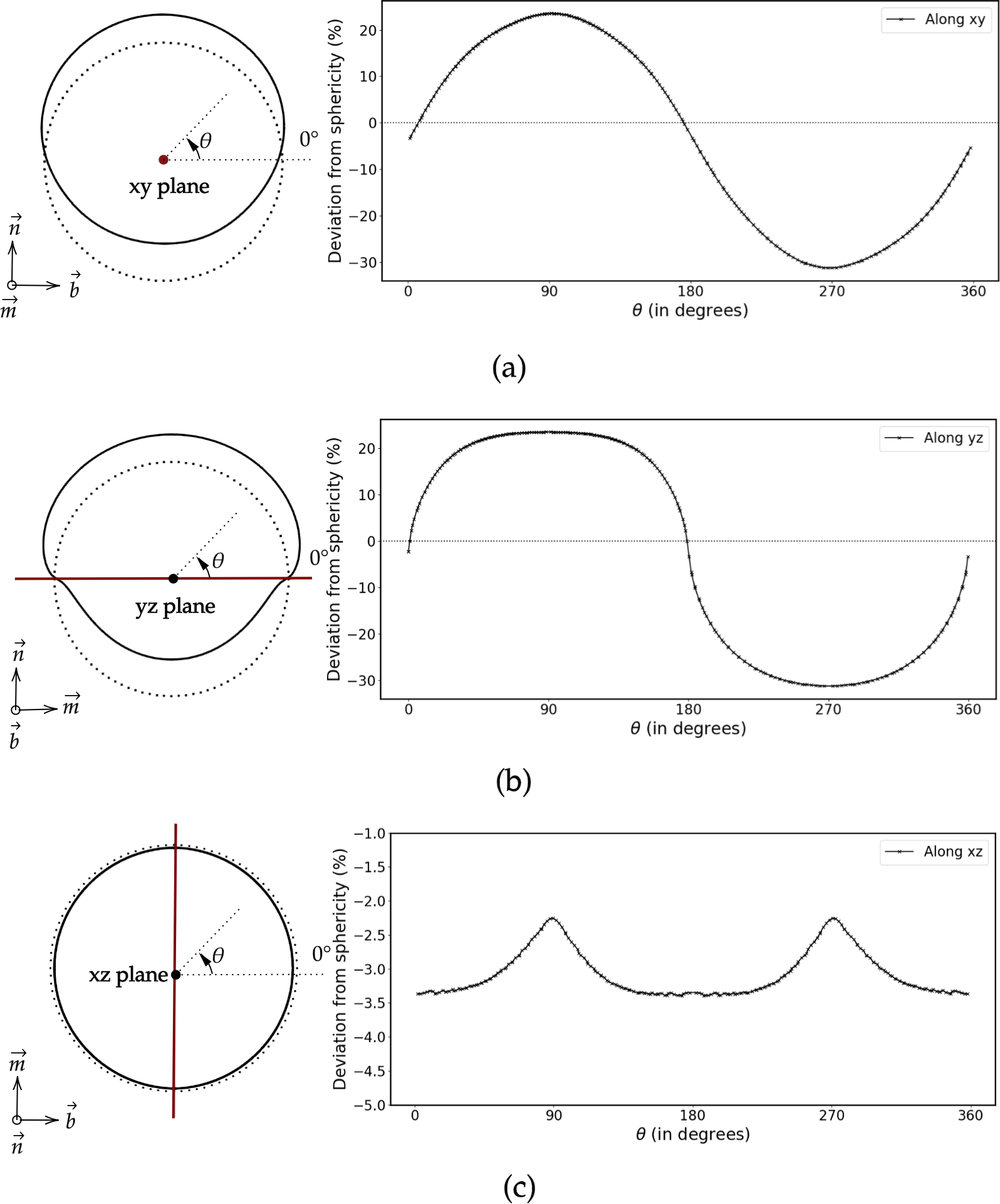}}}\hspace{5pt}
\caption{The \% deviation (from spherical shape) of the morphology of the smaller precipitate $P_S$ connected by an edge dislocation. The deviation is measured by taking the radial distance between the two profiles as a fraction of the radius of the sphere.}
 \label{fig:edge_deviation}
\end{figure}

The screw connecting segment, which only has deviatoric elastic field components (for linear elasticity), does not interact with the purely dilatational strain field of the precipitate. Due to this, the shape of the precipitate is not affected by the elastic interaction. Hence, the morphology of the precipitates with screw connecting dislocation is identical to the morphology in the CCED (a) case, considering only pipe diffusivity of the dislocation.

However, for the edge connecting dislocation, interaction between the dilatational components of the dislocation and the elastic fields of the precipitate results in significant change of shape of the precipitate. The unfavourable elastic fields below the slip plane causes material to flow to above the slip plane, locally along the precipitate surface. Due to this, the precipitates develop a protrusion parallel to the dislocation line above the slip plane, as clearly seen from the $yz$ section of $P_S$ in the second row in Figure~\ref{fig:edge_screw_morphology}. 

For the edge dislocation case, the deviation from sphericity of $P_S$ is calculated by taking the radial distance of the precipitate section from a coincident sphere and is expressed as a fraction of the radius of this sphere, for different angles about the horizontal, as shown in Figure~\ref{fig:edge_deviation}. The deviation is considered positive if the portion of the precipitate is lying outside the sphere and negative otherwise. The movement of material from below 
the slip plane to above is clearly visible in the $xy$ and $yz$ sections. The formation of a protrusion along the dislocation line ($z$-direction) is also observed, indicating a change in the local curvature, which is expected to alter the local equilibrium interfacial composition. 

\begin{figure}[h!]
\centering
\subfloat[]{%
\resizebox*{7cm}{!}{\includegraphics[width=7cm]{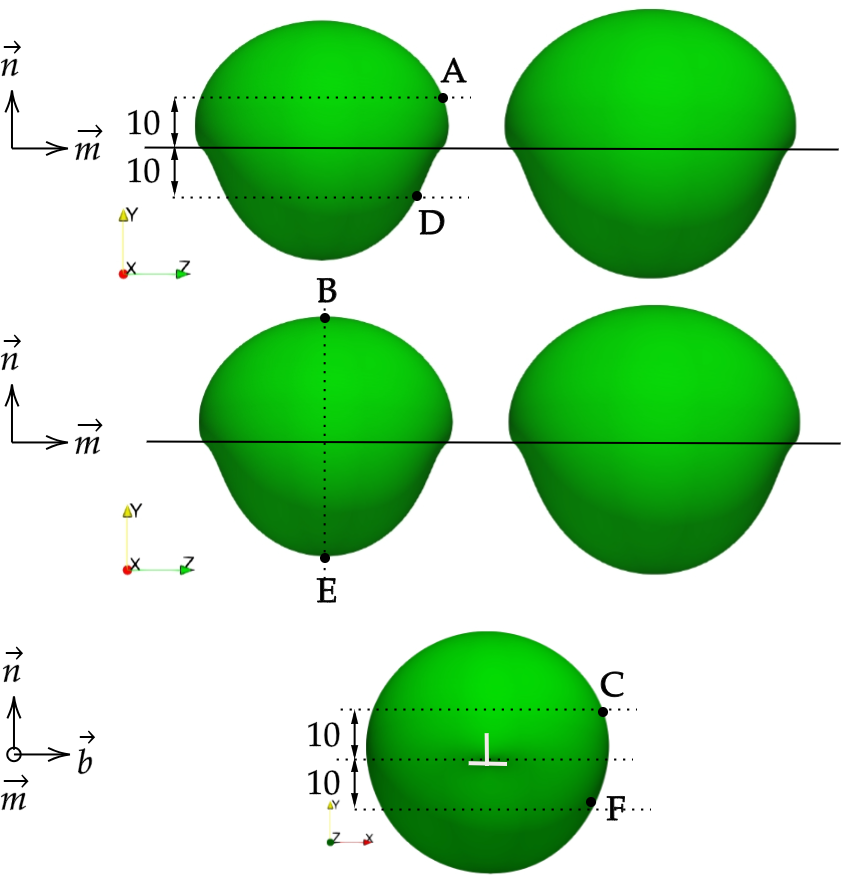}}}\hspace{5pt}
\subfloat[]{%
\resizebox*{7cm}{!}{\includegraphics[width=7cm]{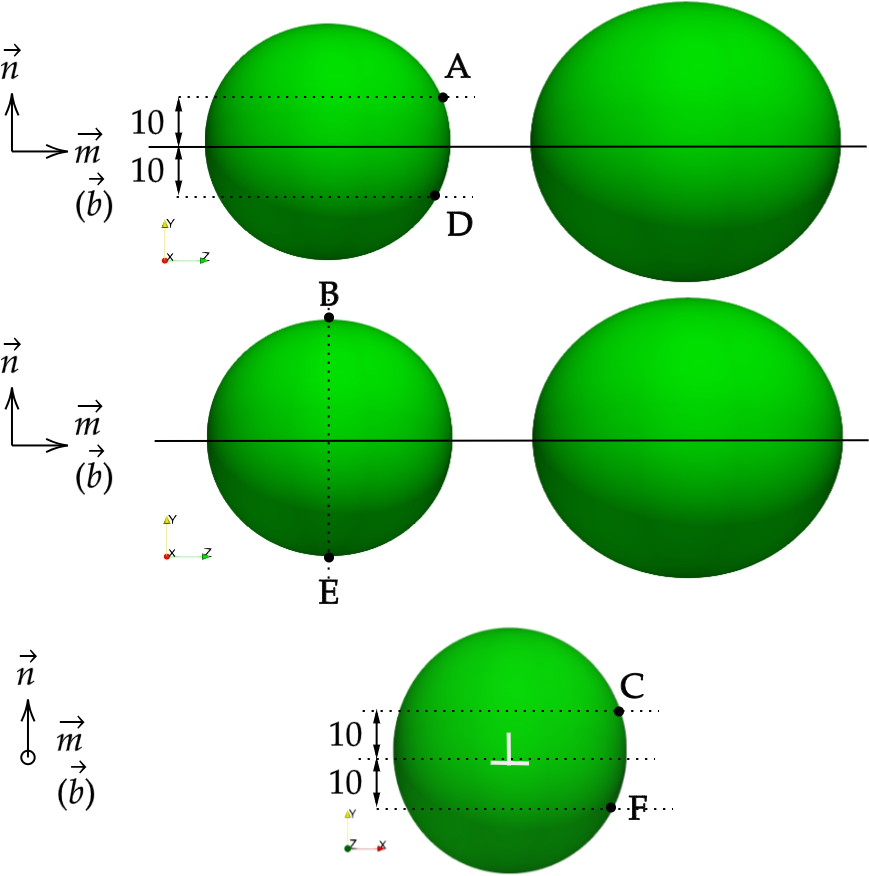}}}\hspace{5pt}
\caption{Points on the surface of precipitate connected by (a) edge and (b) screw dislocations at which $\Delta c^p_{tot}$ is calculated using modified Gibbs-Thomson equation.} \label{fig:AtoF_schematic}
\end{figure}

As there is a significant change in the shape of the precipitate from the initial spherical shape, it is not possible to calculate the $\Delta c^p_{el}$ for the CCED cases by fitting $\Delta c^p_{tot}$ against the mean curvature ($\chi^p$). Since the mean curvature ($\chi^p$) would be different at different points on the surface of the precipitate, we consider six points on the precipitate surface, three above and three below the slip plane, as shown in Figure~\ref{fig:AtoF_schematic}, where the $\Delta c^p_{tot}$ is calculated using the stress and strain fields and the mean curvature of the interface calculated at those points using the parallel surface method~\cite{Nishikawa1998}. These values are then compared with the $\Delta c^p_{tot}$ obtained from the composition profile, thus checking the applicability of the modified Gibbs-Thomson equation in predicting the changed interfacial equilibrium due to the dislocations. As the dislocation eigenstrain is discontinuous along the $y$-direction, the elastic fields calculated by FFT method oscillate due to Gibbs phenomenon. Hence, the stress and strain fields are smoothened by taking a five-point average along the $y$-direction.  

The values of $\Delta c^p_{tot}$ calculated at 
the six points for the two types of dislocations, 
along with the value obtained from the 
composition profiles and the \% relative error is 
tabulated in Table~\ref{tab:deltacp_edge_screw}. 
There is a broad agreement between the values 
obtained by calculations based on the modified 
Gibbs-Thomson equation and the equilibrium 
composition read from the composition profile. 
The difference is of the order of $10\%$ which 
may be attributed to two factors: (a) the error 
in the calculation of mean curvature ($\chi^p$) 
using the parallel surface method 
(PSM)~\cite{Nishikawa1998}, and (b) the 
assumption made while deriving the modified 
Gibbs-Thomson equation that farfield composition 
($c_{f}$) remains a constant, which is not fully 
satisfied in our simulations due to the 
limitations in system size. 

\begin{table}[h!]
\tbl{$\Delta c^p_{tot}$ contributions from 
elastic and interfacial terms calculated by 
substituting the stress and strain fields at 
different points on the precipitate in the 
modified Gibbs-Thomson equation and compared with 
the equilibrium composition from the composition 
profile at the same point. The values for three 
cases: namely (a) without dislocation, (b) screw 
and (c) edge type dislocations are tabulated.}
{\begin{tabular}{lccccc} 
\toprule \\
& \multicolumn{2}{c}{\centering From modified 
G-T equation} & & $\Delta c^p_{tot}$ & \% error 
\\ \cmidrule{2-4} 
 Point & $\Delta c^p_{el}$ (1) & $\Delta c^p_{int}$ (2) & $\Delta c^p_{tot}$ (1)+(2) &  from comp. profile &  \\ \midrule
\textbf{No dislocation} & & & & & \\
A & 0.00554 & 0.01069 & 0.0162 & 0.0150 & 8\\
B & 0.00531 & 0.01109 & 0.0164 & 0.0150 & 9\\
C & 0.00572 & 0.01064 & 0.0164 & 0.0150 & 9\\
D & 0.00552 & 0.01069 & 0.0162 & 0.0150 & 8\\
E & 0.00529 & 0.01109 & 0.0164 & 0.0150 & 9\\
F & 0.00571 & 0.01064 & 0.0164 & 0.0150 & 9\\
\textbf{Screw} & & & & & \\
A & 0.00547  & 0.01075 & 0.0162 & 0.0150 & 8\\
B &  0.00528 & 0.01123 & 0.0165 & 0.0150 & 10\\
C & 0.00574 & 0.01084 & 0.0166 & 0.0150 & 11\\
D & 0.00555 & 0.01075 & 0.0163 & 0.0150 & 9\\
E & 0.00524 & 0.01123 & 0.0165 & 0.0150 & 10\\
F & 0.00575 & 0.01084 & 0.0166 & 0.0150 & 11\\
\textbf{Edge} & & & & & \\
A & -0.000547${^\dagger}$  & 0.01657${^\dagger}$ 
& 0.0160 & 0.0146 & 10\\
B & 0.00427 & 0.01030 & 0.0146 & 0.0133 & 10\\
C & 0.00481 & 0.01150 & 0.0163 & 0.0149 & 9\\
D & 0.01169${^\dagger}$ & 0.00487${^\dagger}$ & 
0.0166 & 0.0154 & 8\\
E & 0.00593 & 0.01220 & 0.0181 & 0.0163 & 11\\
F & 0.00660 & 0.01030 & 0.0169 & 0.0155 & 9\\ 
\bottomrule
\end{tabular}}
\tabnote{$^{\dagger}$ The values marked show an order of magnitude 
difference in the elastic contribution $\Delta 
c^p_{el}$ and its effect on the 
interfacial contribution $\Delta c^p_{int}$ in the case of edge dislocation.}
\label{tab:deltacp_edge_screw}
\end{table}

Another important point is the order of magnitude difference in the $\Delta c^p_{el}$ contribution (marked with $\dagger$ symbol in the table) between the points A and D. This is due to the elastic fields of the dislocation which is equal in magnitude but opposite in direction above and below the slip plane. This difference causes material flow from below to above the slip plane, forming a protrusion above the slip plane along the direction of the dislocation line and hence, a larger mean curvature ($\chi^p$) at A as compared to D. Hence, the elastic fields of the dislocation does not only change the elastic contribution of the modified Gibbs-Thomson equation, but also influences the curvature contribution. 

\subsection{Case CCID: Effect of dislocations in elastically inhomogeneous systems}
\label{CCID_case}
\begin{figure}[h!]
\centering
\subfloat{%
\resizebox*{14cm}{!}{\includegraphics[width=10cm]{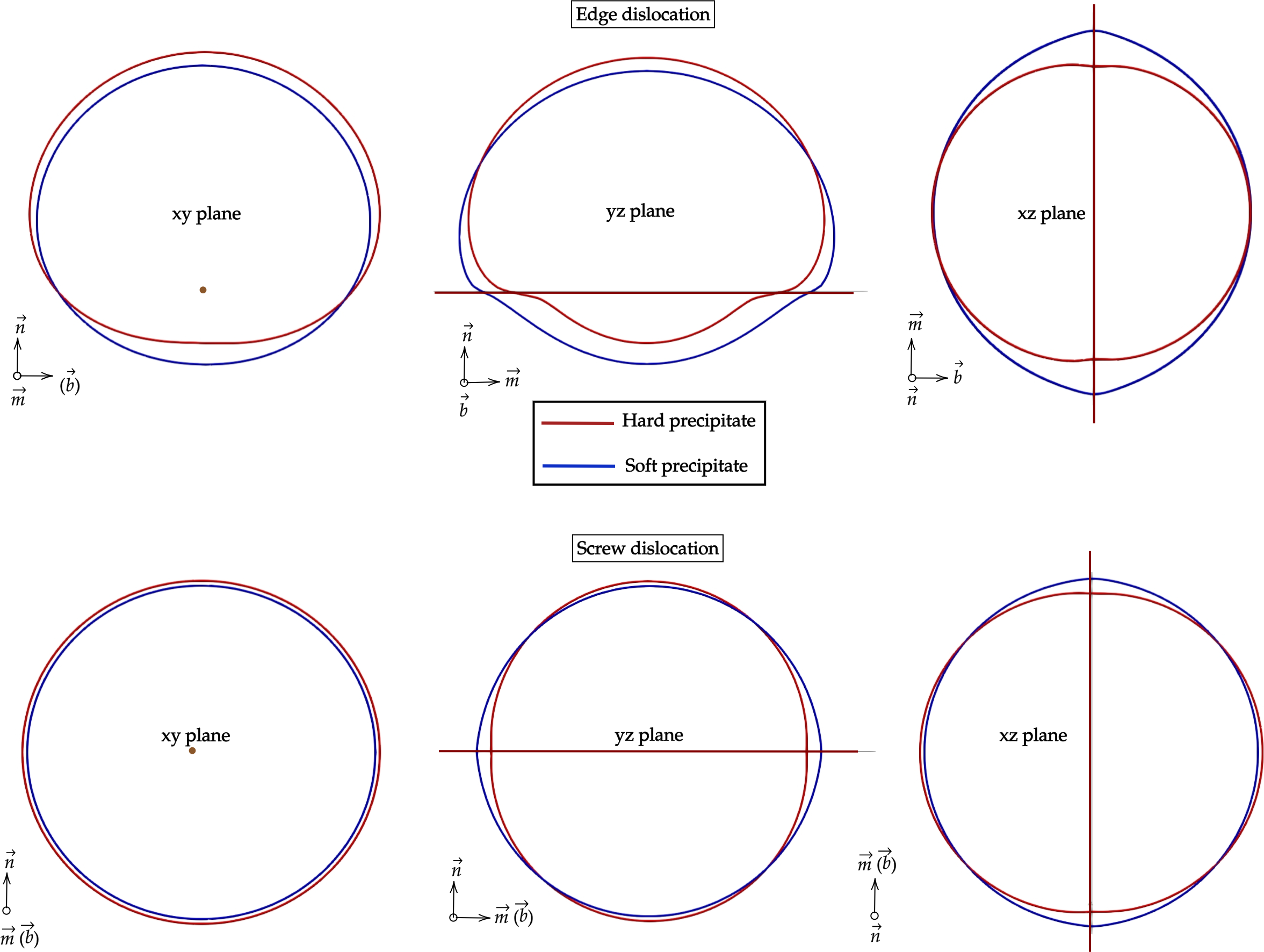}}}\hspace{5pt}
\caption{Morphology of the smaller precipitate 
$P_S$ connected by edge (top row) and screw 
(bottom row) dislocations respectively.} 
\label{fig:hard_soft}
\end{figure}

Next, we consider the effect of elastic inhomogeneity of the system, by considering two sets of simulations with precipitate moduli 20 \% higher ($C^p=1.2C^m$) and 20 \% lower ($C^p=0.8C^m$) compared to the matrix phase, for the edge and screw connecting dislocations. In Figure~\ref{fig:hard_soft}, we show the superimposed $xy$, $yz$ and $xz$ sections of the hard and soft precipitate for the edge and screw connecting dislocations at $R_{\mathrm{eff}} = 28$. There is a marked difference in the shape of the hard and soft precipitates while connected by the same 
dislocation type. In the case of the edge connecting dislocation, the rate of material transport from below to above the slip plane is faster in the case of the hard precipitate, as clearly seen in the $xy$ abd the $yz$ sections. The protrusion formed along the direction of the dislocation line -- due to the material flow from below to above the slip plane -- is more pronounced for the soft precipitate than the hard precipitate, which is evident from the $yz$ and $xz$ views on the first row. Even for the case of screw dislocation, there is a localised shape change near the dislocation line, with a protrusion occurring for the soft precipitate at the point of exit of the dislocation from the precipitate, and a dent on the surface, at the vicinity of the dislocation, for the hard precipitate. 

Unlike the CCED (b) case, there is an interaction of the elastic field of the precipitate with the (purely deviatoric) elastic fields of the screw dislocation, causing a shape change. In an inhomoheneous system, the elastic energy integral can be expanded and the elastic component of the driving force can be obtained by taking its variational derivative as follows~\cite{ArjunThesis}:
\begin{multline}
	\frac{\delta 
		\left[\frac{F^{\mathrm{el}}}{N_V}\right]}{\delta 
		c} = 
	-C^{\mathrm{0\prime}}_{ijkl}\epsilon^c\delta_{ij}\beta^{\prime}(c)(E_{kl}+\epsilon^{*}_{kl}
	- \sum_{s=1}^{N_s}\epsilon^{0d}_{kl}(\eta^s) - 
	\epsilon^{0c}_{kl}(c)) + \\
	\alpha^{\prime}(c)\Delta C^{\prime}_{ijkl}(E_{ij} 
	+ 
	\epsilon^*_{ij} - 
	\sum_{s=1}^{N_s}\epsilon^{0d}_{ij}(\eta^s) - 
	\epsilon^{0c}_{ij}(c)) (E_{kl} + 
	\epsilon^*_{kl} - 
	\sum_{s=1}^{N_s}\epsilon^{0d}_{kl}(\eta^s) - 
	\epsilon^{0c}_{kl}(c))
\label{eq:driving-force}
\end{multline}

where $C^{\mathrm{0\prime}}_{ijkl} = 
\frac{C^{\mathrm{0}}_{ijkl}}{N_V}$, $\Delta 
C^{\prime}_{ijkl} = \frac{\Delta C_{ijkl}}{N_V}$, $\alpha(c)$ and 
$\beta(c)$ are interpolating functions, $E_{ij}$ is the homogeneous (macroscopic)
strain tensor, $\epsilon^*_{ij}$ is the periodic (microscopic) strain, and, 
$\epsilon^{0c}$ and $\epsilon^{0d}$ are the eigenstrains associated 
with the composition and dislocation fields. The first term on the 
right hand side of eq.~(\ref{eq:driving-force}) is identical for 
homogeneous and inhomogeneous systems, with the exception of the 
value of $C^{\mathrm{0\prime}}_{ijkl}$ used. There is no contribution due to the deviatoric elastic fields in the first term, due to the presence of $\delta_{ij}$ 
resulting from the derivative of $\epsilon^{0c}$ (following the 
assumption that precipitate elastic fields are purely dilatational). Note that the sign of this term is also negative. The second term is present only in the case of inhomogeneous systems, and does not reduce to zero for deviatoric fields. Moreover, all variables except for $\Delta 
C^{\prime}_{ijkl}$ are positive. $\Delta C^{\prime}_{ijkl}$ is 
positive (negative) for $C^p>C^m$ ($C^p<C^m$) and hence, the second term will counteract (augment) the first term for the hard (soft) precipitate. Due to this, there is a gradient in the elastic potential which moves material away from the dislocation for the hard precipitate and towards the dislocation line for the soft precipitate. As the  deviatoric fields are highest near the dislocation lines and decreasing away from it, this causes the formation of the protrusion at the surface of the soft precipitate and the depression at the surface of the hard precipitate close to the dislocation, in the case of screw dislocation. The variation in morphology for hard and soft precipitates with edge connecting 
dislocations is also due to their interaction with the deviatoric 
elastic field of the edge connecting dislocation.

Further, we also performed the calculation of the $\Delta c^p_{tot}$ at the points A to F similar to the CCED (with elastic fields and pipe diffusivity) simulations. In this case also, there is a broad agreement of the equilibrium composition values calculated from the modified Gibbs-Thomson equation with that obtained from the composition profile, suggesting that the modified Gibbs-Thomson equation is applicable in this case as well. 

\subsection{Comparison of rates of coarsening}

\begin{figure}[h!]
\centering
\subfloat{%
\resizebox*{8cm}{!}{\includegraphics[width=7cm]{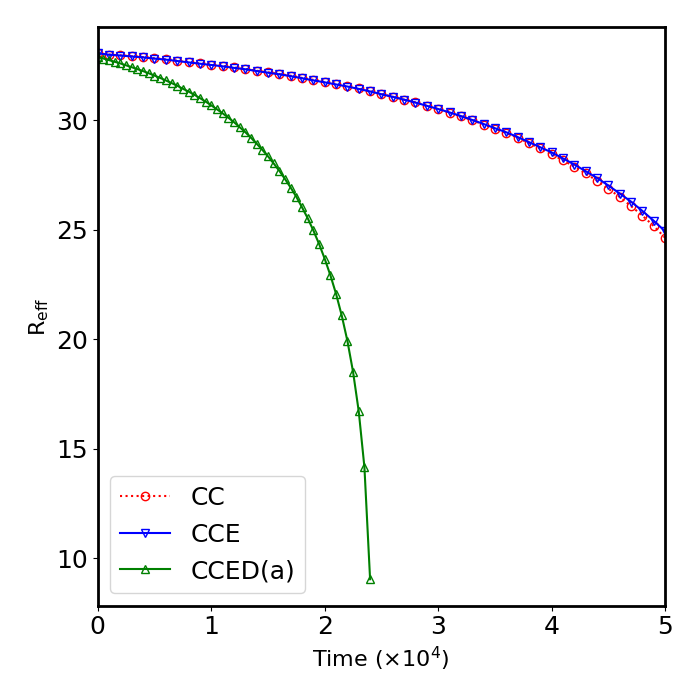}}}\hspace{5pt}
\caption{$R_{\mathrm{eff}}$ as a function of time for CC, CCE and CCED(a) cases. The coarsening rates are almost identical for the CC and CCE cases as the elastic fields of the precipitates are independent of the size. The coarsening rates for the CCED case is faster due to pipe diffusion. The timescale starts at t=0 corresponding to the CCED(a) case, after initial adjustments, and the comparison is made with corresponding timesteps from the CC and CCE cases.}
\label{fig:CC_CCE_CCED}
\end{figure}

Finally, we make a comparison of the coarsening rates of the different simulation cases. In Figure~\ref{fig:CC_CCE_CCED}, we show a comparison of the effective radius $R_{\mathrm{eff}}$ for the smaller precipitate for the CC, CCE and CCED(a) cases, showing the effect of pipe diffusivity on the rate of coarsening. 

The coarsening rates for the CC and the CCE cases are almost identical in our simulations. It has been shown that the elastic fields of a coherent precipitate are independent of the size of the precipitate~\cite{Eshelby1957}. Hence, the elastic contribution to the Gibbs-Thomson effect is the same outside both precipitates, leaving the concentration gradient between them unaffected. Note that they are not exactly identical due to the slightly larger supersaturation of $P_S$ in the CCE case for $c_f = 0.0165$, as shown by our finite size scaling analysis. The faster pipe diffusion along the dislocation accelerates the kinetics for the CCED(a) case.

\begin{figure}[h!]
\centering
\subfloat[]{%
\resizebox*{9.5cm}{!}{\includegraphics[width=9.5cm]{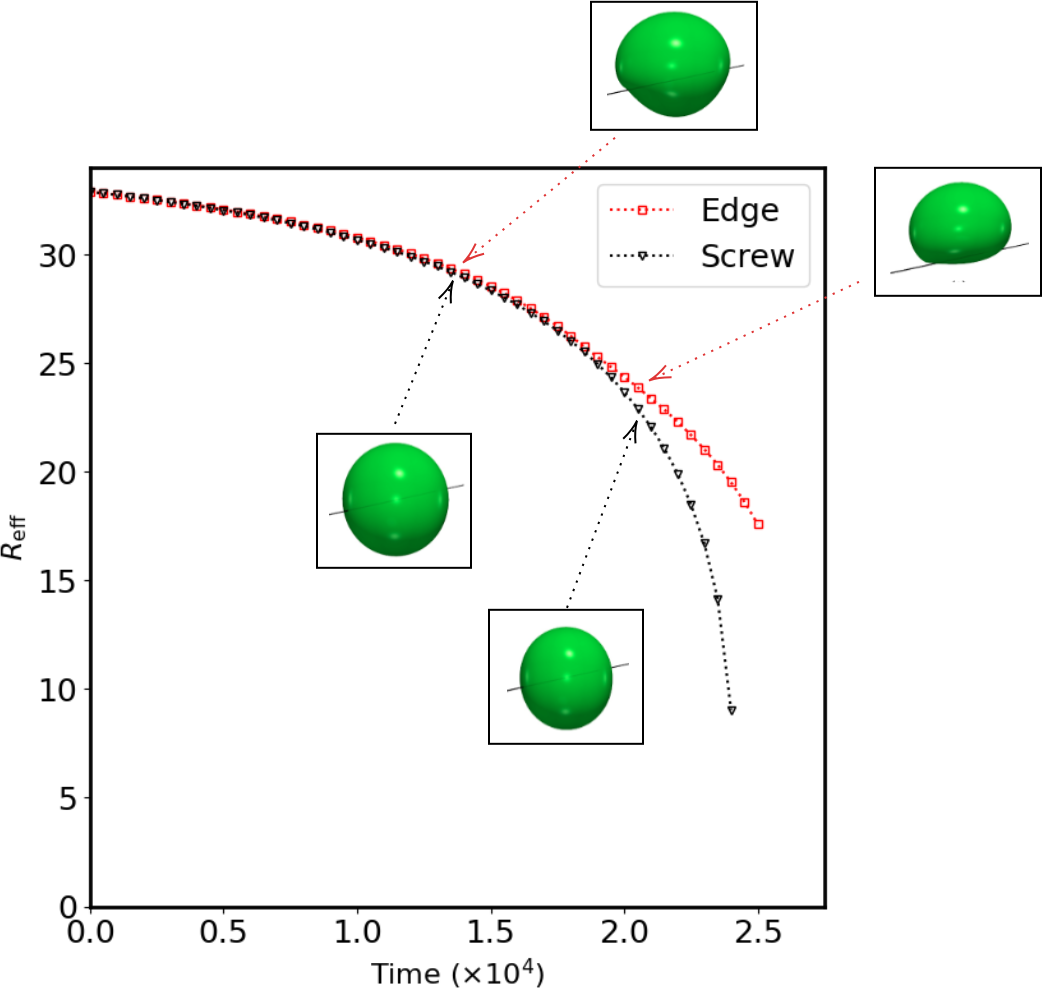}}}\hspace{5pt}
\caption{The evolution effective radius $\mathrm{R}_{\mathrm{eff}}$ for the edge and screw dislocation cases (CCED (b)) along with the morphology of the precipitates shown for t=12500 and t=20000. The insets indicated by red arrows correspond to the edge type connecting dislocation and the black arrows correspond to the screw type connecting dislocation.}
\label{fig:edge_radius_morphology}
\end{figure}

We show the evolution of effective radius for the CCED(b) case in Figure~\ref{fig:edge_radius_morphology}. The kinetics of coarsening is the same for the screw connecting dislocation as compared to the CCED(a) case (with only pipe diffusivity), as the deviatoric strain fields of the screw dislocation do not interact with the dilatational misfit strain fields of the precipitates. Also, the kinetics is slower in the case of the edge dislocation, as shown in Figure~\ref{fig:edge_radius_morphology}, as the solutes in the precipitate find the dislocation elastic fields above the slip plane to be favourable as compared to below the slip plane, setting up a flux from below to above the slip plane. The subsequent shape change due to this flux causes a change in the equilibrium interfacial compositions of the two precipitates, reducing the concentration gradient. This makes the kinetics of coarsening slower in the case of the edge dislocations. 

\begin{figure}[h!]
\centering
\subfloat[]{%
\resizebox*{7cm}{!}{\includegraphics[width=7cm]{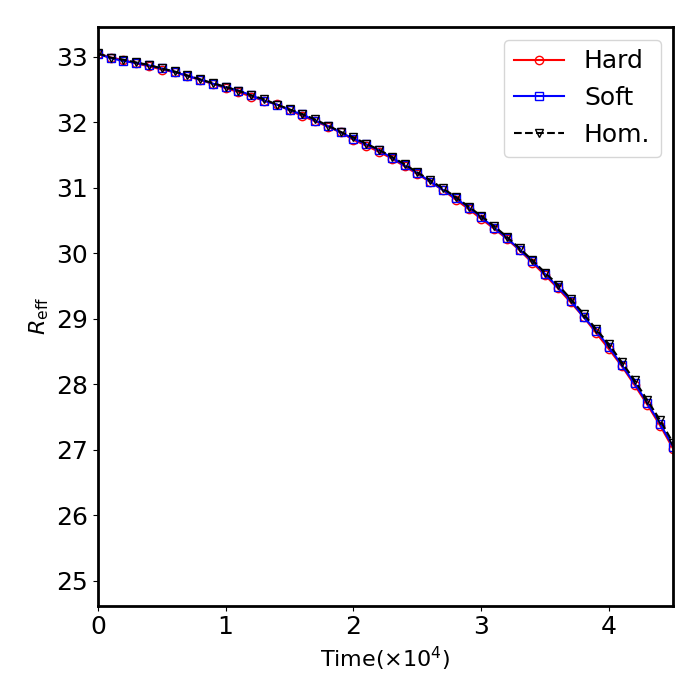}}}\hspace{5pt}
\subfloat[]{%
\resizebox*{7cm}{!}{\includegraphics[width=7cm]{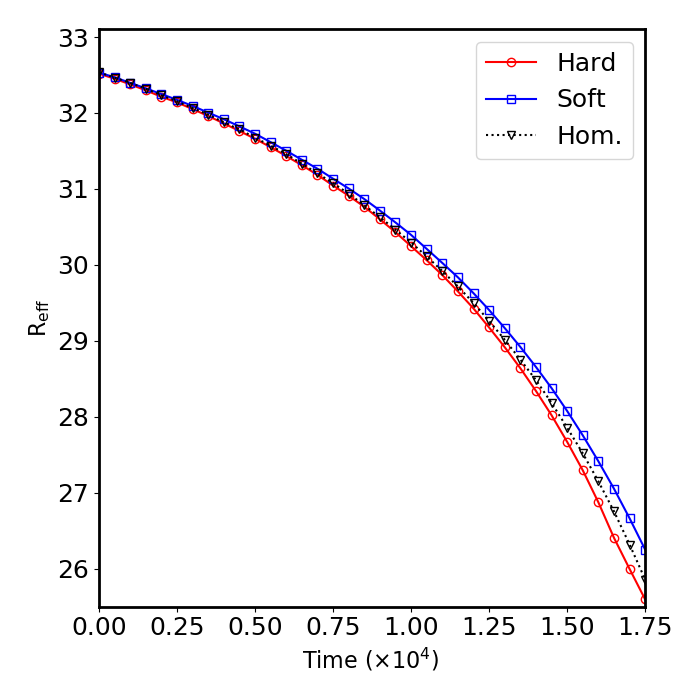}}}\hspace{5pt}
\caption{$R_{\mathrm{eff}}$ as a function of time 
for (a) CCE and CCI (hard and soft precipitate) 
cases (without dislocations and (b) CCED and CCID 
(Hard and soft precipitate) cases (with 
dislocations). The coarsening kinetics is 
identical for the CCE and CCI cases. It is clear 
from (b) that dislocations affect the coarsening 
kinetics differently for systems with hard or 
soft precipitates.} \label{fig:RvsT}
\end{figure}

We show the effect of elastic inhomogeneity on coarsening rates in Figure~\ref{fig:RvsT}. In Figure~\ref{fig:RvsT}(a), we compare the CCE and CCI cases, which are the cases of coherent precipitates without dislocations. 
It is clear that the coarsening rates are unaffected by the modulus ratio between the precipitates and matrix phases. In an elastically inhomogeneous system, even though the elastic fields of a precipitate are a function of its shape, as the smaller and larger precipitates both have a spherical shape, the $\Delta c^p_{el}$ contribution to the interfacial equilibrium composition is the same for $P_S$ and $P_L$. Hence, the composition gradient between the precipitates is only a function of the interfacial curvature of the precipitates for CCE and CCI, giving them identical coarsening kinetics. In the CCI case, we had used $c_f=0.0175$ and $c_f=0.0195$ for the systems with soft and hard precipitates respectively. The choice was made such that the total volume of the precipitates was maintained constant for the CCE and CCI cases throughout the simulation. Hence, unlike the CC and CCE cases, there is no deviation in the coarsening kinetics in this comparison.

In Figure~\ref{fig:RvsT} (b), we show the evolution of $R_{\mathrm{eff}}$ of the smaller precipitate $P_S$ for 
the CCED and CCID case with the edge connecting dislocation. The coarsening kinetics is accelerated for the case of harder precipitates and slowed down for the softer precipitates. As described in Section~\ref{CCID_case}, both the dilatational and the deviatoric fields of the dislocations influence the solute flux in the precipitates. The solute flux is directed towards the dislocation in the case of the soft precipitate and away from the dislocation in the case of a hard precipitate. For the edge connecting dislocation, this creates a more prominent protrusion in the case of the soft precipitate as compared to the hard precipitate. Due to this, the concentration gradient between the precipitates is smaller (larger) in the case of the soft (hard) precipitates, resulting in the slower (faster) kinetics as compared to an elastically homogeneous system.  

\section{Conclusions}

We have studied the effect of dislocations on coarsening of coherent precipitates using a phase
field model that accounts for dislocation strain fields, pipe
diffusion and coherency stresses and strains of solutes and precipitates. 

We are able to capture the effect of pipe diffusion on accelerating the coarsening kinetics. In addition, when we incorporate the coherency effects, the driving force for coarsening is also affected. Specifically, the kinetics of coarsening becomes slower when the precipitates are connected by a dislocation segment of the edge type as compared to a case where we do not account for dislocation elastic fields. This
is because the solutes in the precipitate find the dislocation elastic fields above the slip plane favourable and hence they set up a flux from below the slip plane to above the slip plane. This, in turn, decreases the  equilibrium interfacial compositions outside the two precipitates, resulting in a smaller concentration gradient between them. 

In addition, such fluxes driven by the solute-dislocation interactions alter the morphology of
the coarsening precipitates; specifically,
the nature of the dislocation and the elastic
inhomogeneity play a role in determining the morphology. Since coarsening is driven by curvature, the changes in morphology also affects coarsening rates. 

In elastically inhomogeneous systems, the morphological changes are also affected by another phenomenon. The deviatoric components of the elastic fields do not interact with the 
precipitate fields when the elastic moduli of the system is homogeneous. However, in elastically inhomogeneous systems, with hard and soft precipitates, the interaction with the deviatoric fields of the dislocations cause a material build up / removal around the dislocation. This is true even for the case of the screw dislocation segment, in which the precipitate develops a protrusion (depression) at the point of exit of the dislocation, when the elastic moduli of the precipitate is harder (softer) as compared to the matrix phase.

We see that the modified Gibbs-Thomson equation, originally derived for an isolated precipitates in an infinite matrix, is also broadly applicable in the case of coarsening precipitates, situated at close proximity. Even in systems containing 
precipitates connected by dislocations, the change in the equilibrium interfacial 
composition calculated from modified Gibbs-Thomson equation using dislocation elastic fields from the simulation is close to that obtained from the composition profile.

For the same initial precipitate geometries, the coarsening kinetics remain the same for coherent precipitates as compared to the case where driving force due to interfacial curvature is only considered. In fact, we also note that the coarsening kinetics remain the same for the elastically homogeneous and inhomogeneous systems as well, in the absence of dislocations. However, in presence of dislocations, the elastically inhomogeneous systems behave differently as compared to a homogeneous sytem, with a system containing hard (soft) precipitates undergoing a faster (slower) coarsening kinetics.

Due to computational costs, in this study, we have only studied the coarsening of two coherent precipitates connected by a dislocation. A more realistic simulation of large number of particles would need a parallel implementation of the model.

\section*{Acknowledgement(s)}

We thank (a) Dendrite and Space-Time, IIT Bombay, 
(b) Spinode -- the DST-FIST HPC facility, Dept. 
of ME \& MS, IIT Bombay, and (c) C-DAC, Pune for 
high performance computing facilities. This 
project is funded by Science and Engineering 
Research Board, Dept. of Science and Technology, 
Govt. of India though the research grant 
CRG/2019/005060.

\section*{Disclosure statement}

The authors declare that they have no known competing financial
interests or personal relationships that could 
have appeared to influence the work reported in 
this paper.

\bibliography{references}
\newpage
\section*{Supplementary information}

\subsection*{Calculation of farfield composition 
($c_f$) for the CC case}

For the CC case, the $c_f$ is chosen as the 
equilibrium composition outside the precipitate 
for a precipitate of radius $R=33$, obtained 
using the Gibbs-Thomson equation:
\begin{equation}
\Delta c^m_{int} = \frac{2\gamma  
\chi^p}{(c^p_\infty-c^m_\infty)\Psi_m}
\label{eq:curvature-effectS}
\end{equation}

In our simulations, the choice of parameters 
$A^{ch}=1$ and $\kappa=1$ lead to 
$\gamma=\frac{1}{3}$. As we consider a 
double-well potential with minima at  
$c^p_{\infty}=1.0$ and $c^m_{\infty}=0.0$, 
$c^p_{\infty}-c^m_{\infty}=1.0$ (where 
$c_{\infty}$ is the equilibrium composition for a 
system containing a flat interface.) 
$\Psi_m=\frac{\partial^2 f_0}{\partial c^2}$ is 
the curvature of the double-well potential at the 
matrix composition which is equal to 2 for the 
polynomial used as the double-well potential (and 
also equal to the curvature at the precipitate 
equilibrium composition). Therefore, 
eq.~(\ref{eq:curvature-effectS}) becomes:
\begin{equation}
\Delta c^m_{int} = \frac{  
\chi^p}{3}
\label{eq:curvature-effectSB}
\end{equation}
For the radius of $R=33$, the mean curvature is 
given as: $\chi^p = \frac{1}{33}$. Hence, 
$\Delta c^m_{int}$ is given as:
\begin{equation}
\Delta c^m_{int} = \frac{1}{99} = 0.0101
\end{equation} 
Therefore, the composition outside the 
precipitate $c^m_{R=33}$ is given as:
\begin{equation}
c^m_{R=33} = c^m_{R=\infty} + 0.0101 = 0.0 + 
0.0101 = 0.0101
\end{equation}

\subsection*{Calculation of farfield composition 
($c_f$) for the CCE case}

The farfield composition ($c_f$) for the CCE case 
is determined using the modified 
Gibbs-Thomson equation~\cite{Johnson1987}:
\begin{equation}
\Delta c^m_{tot} = \frac{2\gamma 
\chi^p}{(c^p_\infty-c^m_\infty)\Psi_m} + 
\frac{\sigma^m_{ij}(\epsilon^m_{ij}-\epsilon^p_{ij})
 + 
\frac{1}{2}\left[\sigma^p_{ij}(\epsilon^p_{ij}-\epsilon^c\delta_{ij})
-\sigma^m_{ij}\epsilon^m_{ij}\right]}{(c^p_\infty-c^m_\infty)\Psi_m}
\label{eq:modified_GTS}
\end{equation} 

In addition to the interfacial curvature 
contribution of $\Delta c^m_{int}$ determined for 
the CC case, the equilibrium composition will 
also include an elastic contribution of 
Gibbs-Thomson equation, given by the second term. 
For the dilatational eigenstrain of magnitude 
$\epsilon^c=0.01$ in an elastically homogeneous 
system corresponding to moduli parameters of 
$C_{44} = 140$ and $\nu=0.294$, the stress and 
strain fields in the matrix and precipitate 
phases are obtained using the solution of the 
Eshelby inclusion problem~\cite{Eshelby1957} to 
be:
\begin{equation}
\sigma^m = \left[
\begin{matrix}
-0.855556 & 0.0 & 0.0\\
0.0 & 0.427778 & 0.0 \\
0.0& 0.0 & 0.427778 \\
\end{matrix}
\right]
\end{equation}
\begin{equation}
\sigma^p = \left[
\begin{matrix}
-0.855556 & 0.0 & 0.0\\ 
0.0 & -0.855556 & 0.0\\
0.0 & 0.0 & -0.855556 \\
\end{matrix}
\right]
\end{equation}

\begin{equation}
\epsilon^m = \left[
\begin{matrix}
0.006111 & 0.0 & 0.0\\
0.0 & 0.006111& 0.0 \\
0.0 & 0.0 & 0.006111
\end{matrix}
\right]
\end{equation}

\begin{equation}
\epsilon^p = \left[
\begin{matrix}
-0.0122222 & 0.0 & 0.0\\
0.0 & 0.006111& 0.0 \\
0.0 & 0.0 & 0.006111
\end{matrix}
\right]
\end{equation}
From these values, the eq.~(\ref{eq:modified_GTS}) 
can be written as:
\begin{equation}
\Delta c^m_{tot} = 0.0101 + 0.0064 = 0.0165
\end{equation}

As the equilibrium composition for a system with 
a flat interface ( that is, 
$R=\infty$) is $c^m_{\infty}=0.0$, the 
equilibrium composition outside a precipitate of 
radius R=33 is given as:
\begin{equation}
c^m_{R} = c^m_{\infty} + 0.0165 = 0.0 + 0.0165 = 
0.0165
\end{equation}

The farfield composition $c_f$ for the CCI case 
was chosen as 0.0175 and 0.0195 for the cases 
with soft and hard precipitates, respectively, 
such that the total volume of the precipitates 
remain the same in the CCI and the CCE 
cases during coarsening, in order to compare 
the results from both cases.
\newpage
\subsection*{Comparison of CC and CCE kinetics}

\begin{figure}[h!]
	\centering
	\subfloat[]{\includegraphics[width=6.5cm]{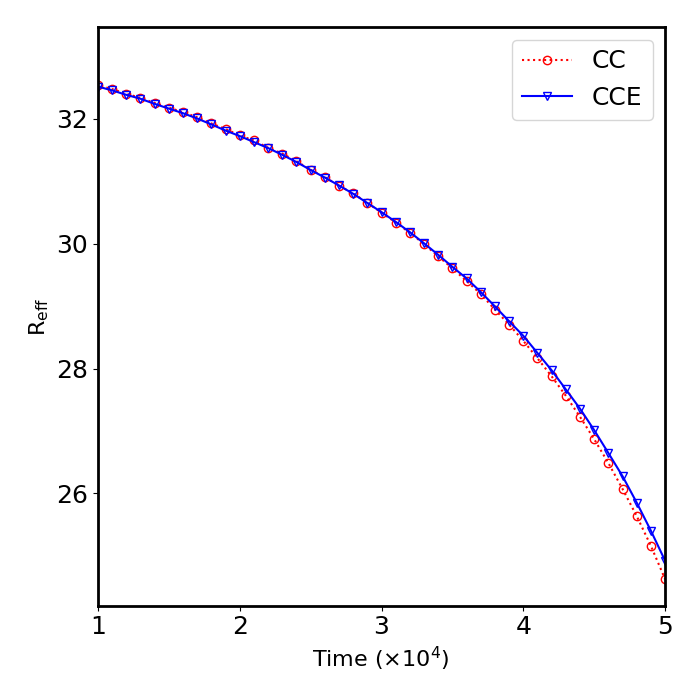}}
	 \hspace{1cm}
	\subfloat[]{\includegraphics[width=6.5cm]{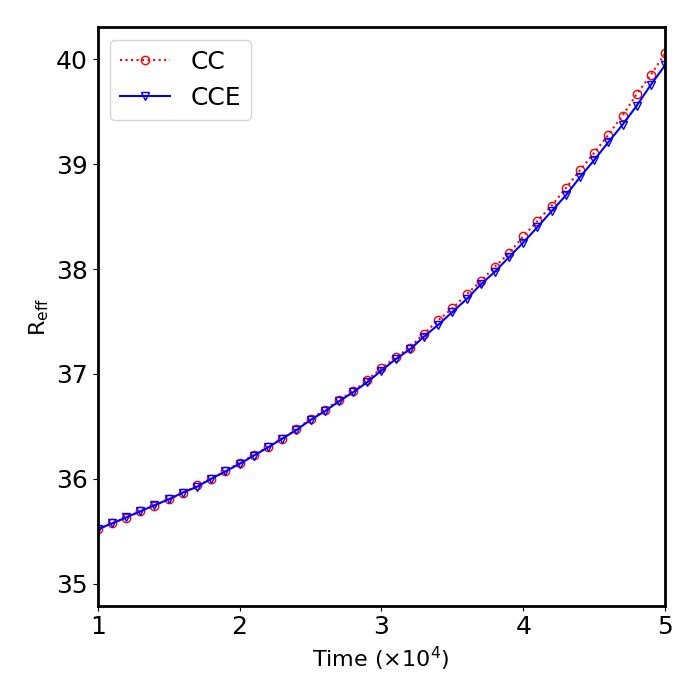}}
	\caption{Effective radii of (a) $P_S$ and (b) 
	$P_L$ as a function of time for the CC and 
	the CCE cases. The initial adjustments 
	upto t=10000 is not shown. The minor 
	difference seen in the evolution of radii is 
	due to the slight supersaturation of $P_S$ in 
	the CCE case.}
	\label{fig:CC_CCE_radii_comparison}
\end{figure} 

\subsection*{Variable mobility due to pipe 
diffusion at dislocations}
\begin{figure}[h!]
	\centering
	\subfloat{\includegraphics[width=13.5cm]{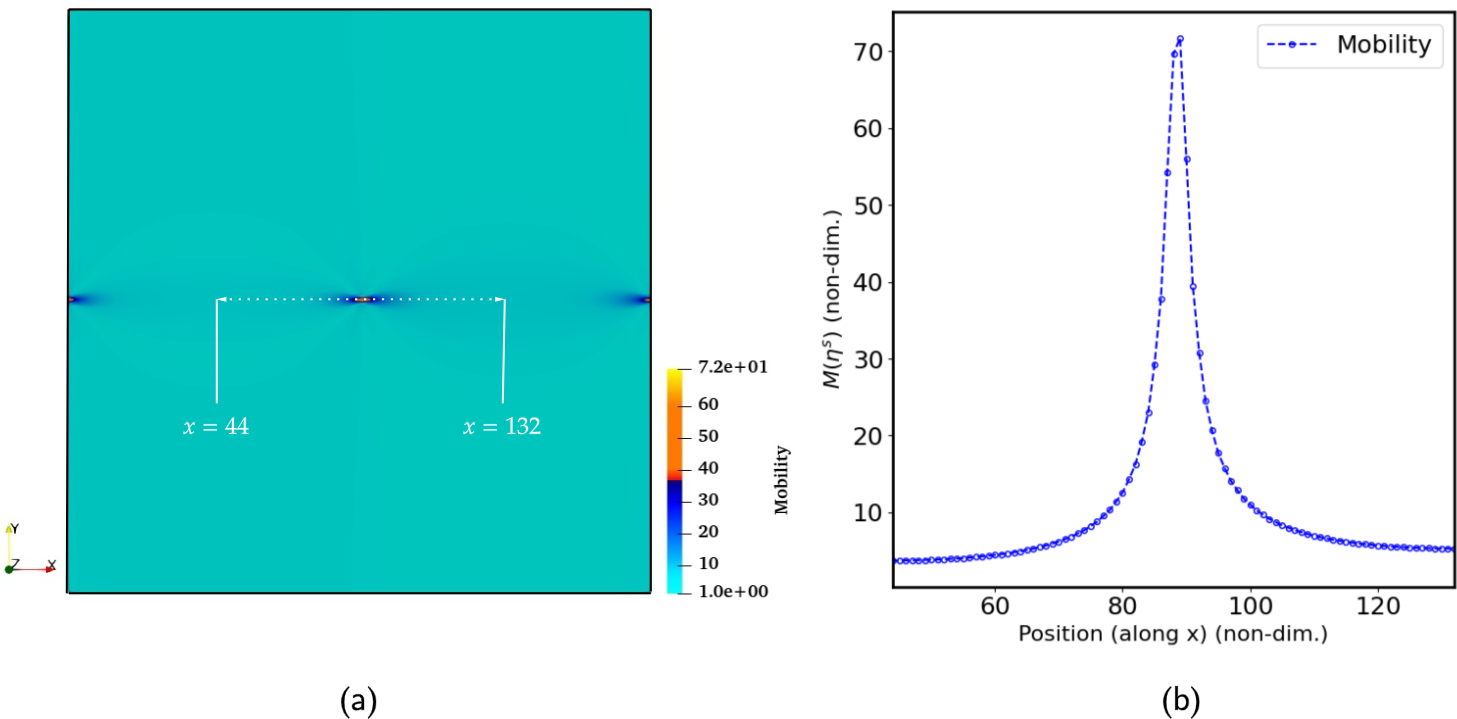}}
	\caption{The mobility map plotted along the 
	direction $z$, perpendicular to the slip 
	plane of the dislocation. In (b), the 
	mobility is plotted along the line shown 
	in (a).}
	\label{fig:CC_CCE_radii_comparisonB}
\end{figure}

\end{document}